\documentclass[reqno,a4paper,final,11pt]{amsart}
\usepackage[british]{babel}
\usepackage{amssymb,amstext,amsthm,eucal}
\usepackage[margin=1in]{geometry}
\usepackage{bbold}
\usepackage{graphicx}
\usepackage{tikz}
\usepackage{array}
\usepackage{tensor} 
\usepackage[utf8]{inputenc}
\usepackage[notref,notcite]{showkeys}
\usepackage[small]{eulervm}
\usepackage{tgpagella}
\usepackage{multicol}
\usepackage{float}
\usepackage{enumerate}
\usepackage{longtable,tabu}
\usepackage[section]{placeins}
\usepackage{hyperref}
\hypersetup{%
  pdftitle   = {Symmetric backgrounds of type IIB supergravity},
  pdfkeywords = {supergravity, Killing spinors, isometries,
    homogeneity, supersymmetry, symmetric spaces, type IIB},
  pdfauthor  = {José Figueroa-O'Farrill, Noel Hustler},
  pdfcreator = {\LaTeX\ with package \flqq hyperref\frqq}
}
\newcommand{\half}{\tfrac{1}{2}}

\newcommand{\fg}{\mathfrak{g}}
\newcommand{\fh}{\mathfrak{h}}

\newcommand{\fm}{\mathfrak{m}}

\newcommand{\fsl}{\mathfrak{sl}}
\newcommand{\fso}{\mathfrak{so}}
\newcommand{\fsp}{\mathfrak{sp}}

\newcommand{\fsu}{\mathfrak{su}}
\newcommand{\fu}{\mathfrak{u}}

\newcommand{\SO}{\mathrm{SO}}

\ifx\Sp\UnDeFiNeD
\newcommand{\Sp}{\mathrm{Sp}}
\else
\renewcommand{\Sp}{\mathrm{Sp}}
\fi

\newcommand{\SL}{\mathrm{SL}}
\newcommand{\SU}{\mathrm{SU}}

\newcommand{\RR}{\mathbb{R}}
\newcommand{\CC}{\mathbb{C}}
\newcommand{\HH}{\mathbb{H}}

\newcommand{\ZZ}{\mathbb{Z}}

\newcommand{\eL}{\mathcal{L}}

\DeclareMathOperator{\CW}{CW}

\DeclareMathOperator{\dS}{dS}
\DeclareMathOperator{\AdS}{AdS}

\DeclareMathOperator{\Ric}{Ric}

\DeclareMathOperator{\SLAG}{SLAG}
\DeclareMathOperator{\ASSOC}{ASSOC}

\newcommand{\CP}{\mathbb{C}\text{P}}

\newcommand{\HP}{\mathbb{H}\text{P}}
\definecolor{orange}{rgb}{0.9,0.45,0}
\newcommand{\MUNCH}[1]{\relax}

\allowdisplaybreaks[1]
\newcommand{\inprod}[2]{\langle #1, #2 \rangle}
\newcommand{\inint}[3]{\inprod{\iota_{#1} #3}{\iota_{#2} #3}}

\newcommand{\nrmsq}[1]{\lvert #1 \rvert^2 }
\newcommand{\then}{\hspace{5pt}\Rightarrow\hspace{5pt}}

\newcommand{\fH}{H^{(3)}}
\newcommand{\fGa}{F^{(1)}}
\newcommand{\fG}{F^{(3)}}
\newcommand{\fF}{F^{(5)}}
\begin{document}
%
\numberwithin{equation}{section}
\usetikzlibrary{arrows,shapes,calc,decorations.pathreplacing,decorations.pathmorphing,positioning}
\setcounter{secnumdepth}{5}
\tikzstyle{every picture}+=[remember picture]
\title[Symmetric IIB backgrounds]{Symmetric backgrounds of type IIB supergravity}
\author[Figueroa-O'Farrill]{José Figueroa-O'Farrill}
\author[Hustler]{Noel Hustler}
\address{Maxwell and Tait Institutes, School of Mathematics, University of Edinburgh}
\thanks{EMPG-12-17}
\begin{abstract}
  In this paper we study homogeneous backgrounds of type IIB supergravity where the underlying geometry is that of a symmetric space.  We determine which ten-dimensional lorentzian symmetric spaces (up to local isometry) admit such backgrounds and in about two thirds of the cases we determine fully their moduli space.
\end{abstract}
\maketitle
\tableofcontents

\section{Introduction}
\label{sec:introduction}

One motivation for studying homogeneous supergravity backgrounds comes from the homogeneity conjecture \cite{FMPHom,EHJGMHom}, reviewed in \cite{JMF-HC-Lecs}, and now a theorem \cite{HomogThm} for 10- and 11-dimensional supergravities, which states that a supergravity background preserving more than half of the supersymmetry is (locally) homogeneous.  Local homogeneity says that there is a basis for the tangent space at every point consisting of Killing vectors which preserve (up to gauge transformations where relevant) all the bosonic fields in the background.  The strong version of the conjecture, which has now been proved for 10- and 11-dimensional supergravities, says that those Killing vectors are actually constructed out of the Killing spinors of the background.

Although this result simplifies the classification of highly supersymmetric backgrounds, it still leaves the nontrivial problem of classifying homogeneous backgrounds.  In a recent paper \cite{FigueroaO'Farrill:2011fj} a first step is made in the classification of homogeneous eleven-dimensional supergravity backgrounds by considering those homogeneous backgrounds where the underlying geometry is that of a symmetric space.  The purpose of the present paper is to do the same for type IIB supergravity backgrounds.   The work follows closely the approach of \cite{FigueroaO'Farrill:2011fj}, to which we refer the reader for much of the underlying motivation and notation concerning lorentzian symmetric spaces.  In a forthcoming paper \cite{FOUM2} a classification of homogeneous M2 duals is presented, which shows that it is feasible to dial up a (semisimple) Lie group $G$ and classify supergravity backgrounds which are homogeneous under the action of $G$.

This paper is organised as follows.  In Section \ref{sec:type-iib-supergr} we will review type IIB supergravity, its field equations and basic symmetries: $\SL(2,\RR)$ duality and its invariance under the homothetic action of $\RR^+$.

In Section \ref{sec:symm-backgr} we will specialise to symmetric backgrounds.  We will first discuss homogeneous backgrounds and introduce the notion of a \emph{strongly homogeneous} background, as a homogeneous background where in addition the axion is constant.  We remark that backgrounds preserving more than half of the supersymmetry are strongly homogeneous and that strong homogeneity is preserved under $\SL(2,\RR)$ duality.  In addition we show that a homogeneous background with underlying geometry $G/H$ is automatically strongly homogeneous if the Lie algebra $\fg$ of $G$ obeys $[\fg,\fg]=\fg$.  In this paper we will study symmetric backgrounds which are not necessary strongly symmetric; although in some cases the results simplify for  strongly symmetric backgrounds and we will mention it when it occurs.  Section \ref{sec:symm-backgr} continues with a list of ten-dimensional lorentzian symmetric spaces, after which we describe the methodology we follow as well as some useful observations and the basic notation we will adhere to in the rest of the paper.

In Section~\ref{sec:analys-spec-cases} we analyse some special cases which are easily dealt with using general arguments instead of detailed calculations.  These are the geometries where the indecomposable lorentzian factor $M_0$ in equation \eqref{eq:decomposition} is de~Sitter, one-dimensional or a Cahen--Wallach space.  This leaves those geometries where $M_0$ is an anti-de~Sitter space.  Those will be studied in detail in Section \ref{sec:analys-rema-spac}, but not without quickly dealing with those geometries where either $M_0$ or one of the $M_i$ is of sufficiently high dimension.

In Section~\ref{sec:analys-rema-spac} we come to the bulk of the detailed results in the paper.  We study symmetric backgrounds with underlying geometry $\AdS_d \times K^{10-d}$ for $2 \leq d \leq 6$ in the order of decreasing $d$.  We discuss the resulting polynomial equations for the parameters which describe the fluxes and in most cases we solve for the moduli space exactly; although as $d$ decreases we face increasingly complicated polynomial systems which we have been thus far unable to solve fully.  In all cases we can however exhibit some some exact solutions, whose existence in many cases was first gleaned from a numerical approach based on crude optimization techniques briefly described in Section \ref{sec:polynomial-systems}.

Finally in Section \ref{sec:summary} we offer some conclusions and summarise the backgrounds found in the paper.  The backgrounds are summarised in Tables~\ref{tab:AdSModuli} and \ref{tab:AdSUnknown}.  The paper ends with an appendix listing geometries which were shown as a result of explicit calculations not to admit symmetric backgrounds, and giving the details of the inadmissibility of several of the trickier geometries.

\section*{Acknowledgments}

This work was supported in part by the grant ST/J000329/1 ``Particle Theory at the Tait Institute'' from the UK Science and Technology Facilities Council.  In addition, NH is supported by an EPSRC studentship.

\section{Type IIB supergravity}
\label{sec:type-iib-supergr}

Type IIB supergravity \cite{SchwarzIIB,SchwarzWestIIB,HoweWestIIB} is the effective field theory of the type IIB superstring.  It is the unique $N=2$, $d=10$ chiral supergravity theory with 32 supercharges and cannot be constructed as a Kaluza--Klein reduction of eleven-dimensional supergravity, but can be related to the non-chiral type IIA theory through T-duality.  The bosonic field content of the theory is: a ten-dimensional lorentzian metric $g$, the dilaton $\phi$, the axion $C^{(0)}$, the R-R gauge potentials $C^{(2)}$ and $C^{(4)}$, and the NS-NS gauge potential $B^{(2)}$.

\subsection{Action and field equations}
\label{sec:acti-field-equat}

There is no covariant action for type IIB supergravity because the theory has a self-dual field strength ($G^{(5)}$).  However, a non-self-dual (NSD) action can be constructed that on variation yields the correct field equations when supplemented with the self-duality condition as an additional field equation.  The bosonic NSD action (in the string frame) is given by
\begin{multline}
  S_{NSD} =  \int \left\{ e^{-2 \phi} \left(R + 4 \nrmsq{d \phi} - \tfrac{1}{2} \nrmsq{H^{(3)}} \right) - \tfrac{1}{2} \left( \nrmsq{G^{(1)}} + \nrmsq{G^{(3)}} + \tfrac{1}{2} \nrmsq{G^{(5)}} \right) \right\} dvol\\
 - \tfrac12 \int C^{(4)} \wedge H^{(3)} \wedge d C^{(2)}~,
\end{multline}
where $R$ is the scalar curvature of $g$, $dvol$ is the signed volume element, and we have introduced the field strengths
\begin{equation}
  \begin{aligned}[m]
    G^{(1)} &= d C^{(0)}\\
    G^{(3)} &= d C^{(2)} - C^{(0)} H^{(3)}\\
    H^{(3)} &= d B^{(2)}\\
    G^{(5)} &= d C^{(4)} - \frac{1}{2} d B^{(2)} \wedge C^{(2)} +
    \frac{1}{2} d C^{(2)} \wedge B^{(2)}~.
  \end{aligned}
\end{equation}
We will not list the gauge transformations here but note that these field strengths are gauge-invariant.  In addition, the inner product on differential forms is defined by
\begin{equation}
	\inprod{X}{Y} dvol = X \wedge \star Y ~,
\end{equation}
and the corresponding (indefinite) norm is
\begin{equation}
	\nrmsq{X} = \inprod{X}{X}~.
\end{equation}

Varying this action with respect to the field potentials and metric yields the following field equations, where we have added the 5-form self-duality equation by hand:
\begin{equation}
  \label{eq:IIBEoM}
  \begin{aligned}[m]
    \Delta \phi &= \tfrac{1}{16} \nrmsq{H^{(3)}} - \tfrac{1}{16} e^{2 \phi} \nrmsq{G^{(3)}} - \tfrac{1}{8} e^{2 \phi} \nrmsq{G^{(1)}}\\
    d \star G^{(1)} &= -H^{(3)} \wedge \star G^{(3)}\\
    d \star G^{(3)} &= -H^{(3)} \wedge G^{(5)}\\
    d \star H^{(3)} &= e^{2 \phi} G^{(3)} \wedge G^{(5)}\\
    d \star G^{(5)} &= H^{(3)} \wedge G^{(3)}\\
    G^{(5)} &= \star G^{(5)}\\
    \Ric(X,Y) &= -4 (X\phi)(Y\phi) + \tfrac12 e^{2 \phi} G^{(1)}(X) G^{(1)}(Y) + \tfrac{1}{2} \inint{X}{Y}{H^{(3)}} + \tfrac12 e^{2 \phi} \inint{X}{Y}{G^{(3)}}\\
      & \qquad {} + \tfrac14 e^{2\phi} \inint{X}{Y}{G^{(5)}} - \tfrac{1}{8}g(X,Y)
      \nrmsq{H^{(3)}} - \tfrac{1}{8} e^{2\phi} g(X,Y) \nrmsq{G^{(3)}}~,
  \end{aligned}
\end{equation}
where $\Ric$ stands for the Ricci tensor.

\subsection{\texorpdfstring{$\SL(2,\RR)$}{SL(2,\mathbb{R})} symmetry}
\label{sec:sl2R}

The Type IIB NSD action exhibits a global $SL(2,\mathbb{R})$ symmetry \cite{SchwarzWestIIB} under which (and in the Einstein frame) $g$ and $C^{(4)}$ are inert, $C^{(2)}$ and $B^{(2)}$ transform as a doublet and the axi-dilaton $\tau = C^{(0)} + i e^{-\phi}$ transforms via fractional linear transformations in the upper-half plane.  Explicitly, for a group element
\begin{equation}
	\begin{pmatrix} a & b\\ c & d \end{pmatrix} \in SL(2,\mathbb{R})~,
\end{equation}
the transformed $\tau$ and $(B^{(2)},C^{(2)})$ are given by
\begin{equation}
	\tau^{\prime} = \frac{a \tau + b}{c \tau + d} \qquad\text{and}\qquad
        \begin{pmatrix}
		(B^{(2)})^{\prime}\\
		(C^{(2)})^{\prime}
        \end{pmatrix} = 
        \begin{pmatrix}
		a & b\\
		c & d
        \end{pmatrix}
        \begin{pmatrix}
          B^{(2)}\\
          C^{(2)}
        \end{pmatrix}~.
\end{equation}
We note that the type IIB string theory preserves the $SL(2,\mathbb{Z})$ subgroup of this symmetry corresponding to those matrices where $a,b,c,d \in \ZZ$.

\subsection{Homothety invariance of the field equations}
\label{sec:homoth-invar-field}

The field equations \eqref{eq:IIBEoM} are invariant under the homothetic action of $\RR^+$ given by
\begin{equation}
  \label{eq:homothety}
  \left(g,\phi,G^{(1)},G^{(3)},G^{(5)},H^{(3)}\right) \mapsto \left(e^{2t}g,\phi,G^{(1)},e^{2t}G^{(3)},e^{4t}G^{(5)},e^{2t}H^{(3)}\right)~,
\end{equation}
where $t \in \RR$.  Indeed, under $g \mapsto e^{2t} g$, the Levi-Civita connection is inert, consisting as it does of terms of the form $g^{-1}dg$.  This means that the $(3,1)$ Riemann curvature tensor is also inert, and so is any of its contractions, such as the Ricci tensor.  Also, the Hodge $\star$ acting on $p$-forms scales like $e^{(10-2p)t}$ under $g \mapsto e^{2t}g$.  This is enough to check that all equations in \eqref{eq:IIBEoM} scale homogeneously with degrees $-2$, $8$, $6$, $6$, $4$, $4$ and $0$, respectively.

\section{Symmetric backgrounds}
\label{sec:symm-backgr}

We now specialise to symmetric backgrounds.  Symmetric backgrounds are special cases of homogeneous backgrounds, so we discuss these first.

\subsection{Homogeneous backgrounds}
\label{sec:homog-backgr}

These are backgrounds where the underlying geometry is that of a homogeneous lorentzian manifold, so that there is a Lie group $G$ acting transitively via isometries.  In addition we demand that all bosonic fields are $G$-invariant.  In a theory like type IIB supergravity, which has a formulation in terms of a gauge theory, one has to allow for the possibility that $G$ leaves invariant the fields only up to gauge transformations.  However this is easy to implement by demanding $G$-invariance of the gauge-invariant field strengths.  This means that we will take all the bosonic fields $g$, $\phi$, $G^{(1)}$, $G^{(3)}$, $G^{(5)}$ and $H^{(3)}$ to be $G$-invariant.  In particular this says that $\phi$ is constant.  In many cases, it will also mean that $C^{(0)}$ is also constant, whence so is the axi-dilaton $\tau$, which may be then transformed to $\tau = i$ via an $\SL(2,\RR)$ duality transformation.  The subgroup of $\SL(2,\RR)$ which fixes $i$ is precisely $\SO(2)$, whence homogeneous backgrounds where $\tau=i$ and $G^{(3)}$ (and hence $\fH$) are nonzero still come in families parametrised by an angle, corresponding to the orbit of any one of these backgrounds under the action of $\SO(2)$.

It is convenient to introduce the notion of a \textbf{strongly homogeneous} background to be a homogeneous background where in addition $C^{(0)}$ is constant.  In \cite{EmilyIIB} (see also \cite{EHJGMHom}) it is shown that the Killing vectors constructed out of Killing spinors preserve $C^{(0)}$, whence the recent proof \cite{HomogThm} of the strong homogeneity conjecture implies that any type IIB backgrounds preserving more than one half of the supersymmetry are strongly homogeneous.

It should be remarked that whereas the $\SL(2,\RR)$-dual of a strongly homogeneous background is again strongly homogeneous, this is not the case for homogeneous backgrounds in general.  Indeed, for a general homogeneous background, the dilaton is constant, but under a general $\SL(2,\RR)$ transformation the dilaton transforms into a function of the dilaton and the axion.  If the axion is not constant, then the new transformed dilaton will not be either and hence the transformed background cannot be homogeneous.

A homogeneous background with underlying geometry $G/H$ is forced to be strongly homogeneous if it does not admit any $G$-invariant one-forms.  In addition it might be forced to be strongly homogeneous depending on $G$ or, more precisely, on the the Lie algebra $\fg$ of $G$, as we now explain.

If $X \in \fg$ we will let $\hat X$ denote the corresponding Killing vector on $G/H$ and $\eL_{\hat X}$ the corresponding Lie derivative.  Since $G^{(1)} = d C^{(0)}$ is $G$-invariant, $\eL_{\hat X} G^{(1)} = 0$, which is equivalent to $d \eL_{\hat X} C^{(0)} = 0$ or, in other words, that $\eL_{\hat X} C^{(0)}$ is a constant.  Being linear in $X$, the assignment $X \mapsto \eL_{\hat X} C^{(0)}$ defines a linear map $\alpha : \fg \to \RR$, which we claim to be a cocycle.  Indeed, if $X,Y \in \fg$, then
\begin{equation}
  0 = [\eL_{\hat X}, \eL_{\hat Y}] C^{(0)} = \eL_{[\hat X,\hat Y]} C^{(0)} = \eL_{\widehat{[X,Y]}} C^{(0)} = \alpha([X,Y])~.
\end{equation}
In other words, $\alpha$ annihilates the first derived ideal $[\fg,\fg]$ and hence it defines an element in the first cohomology group $H^1(\fg,\RR)$ of $\fg$ with values in the trivial module $\RR$.  The element $\alpha$ is the obstruction to being able to take $C^{(0)}$ constant.

Lie algebras for which $[\fg,\fg]=\fg$ are called \emph{perfect}.  Semisimple Lie algebras are perfect, for instance.  If $\fg$ is perfect, so that $H^1(\fg,\RR)=0$, then $\alpha=0$ and hence $C^{(0)}$ is constant.  However the Lie algebra of the transvection group of a lorentzian symmetric space is not necessarily perfect.  If the symmetric space has flat directions or if the indecomposable lorentzian factor is a Cahen--Wallach spacetime, the condition $H^1(\fg,\RR)=0$ is not obeyed.  In those cases we also have invariant one-forms in the spacetime and hence $G^{(1)}$ need not vanish.

Using the fact that the dilaton $\phi$ is constant for a homogeneous background, we may actually eliminate it from the field equations \eqref{eq:IIBEoM} for a homogeneous background by introducing $F^{(i)} := e^{\phi} G^{(i)}$, for $i=1,3,5$.  Indeed, the equations become
\begin{equation}
  \label{eq:IIBEoMhom}
  \begin{aligned}[m]
    \nrmsq{H^{(3)}} &= \nrmsq{F^{(3)}} + 2 \nrmsq{F^{(1)}} \\
    d \star F^{(1)} &= \star F^{(3)} \wedge H^{(3)} \\
    d \star F^{(3)} &= F^{(5)} \wedge H^{(3)} \\
    d \star H^{(3)} &= F^{(3)} \wedge F^{(5)} \\
    d \star F^{(5)} &= H^{(3)} \wedge F^{(3)} \\
    F^{(5)} &= \star F^{(5)}\\
    \Ric(X,Y) &= \tfrac12 F^{(1)}(X) F^{(1)}(Y) + \tfrac{1}{2} \inint{X}{Y}{H^{(3)}} + \tfrac12 \inint{X}{Y}{F^{(3)}}\\
      & \qquad {} + \tfrac14 \inint{X}{Y}{F^{(5)}} - \tfrac{1}{8}g(X,Y)
      \nrmsq{H^{(3)}} - \tfrac{1}{8} g(X,Y) \nrmsq{F^{(3)}}~.
  \end{aligned}
\end{equation}
Of course, for a strongly homogeneous background, we have in addition that $F^{(1)}=0$.

\subsection{Symmetric backgrounds}
\label{sec:symmetric-backgrounds}

A (strongly) homogeneous background is said to be \textbf{(strongly) symmetric} if the underlying homogeneous manifold is a lorentzian symmetric space.  These have been discussed in detail in \cite{FigueroaO'Farrill:2011fj} to where we direct the reader for the notation and relevant notions.  One important property of symmetric spaces is that invariant forms are parallel relative to the Levi-Civita connection, whence in particular they are closed and coclosed.  This further simplifies the IIB field equations:
\begin{equation}
  \label{eq:IIBEoMsym}
  \begin{aligned}[m]
    \nrmsq{H^{(3)}} &= \nrmsq{F^{(3)}} + 2 \nrmsq{F^{(1)}}\\
    0 &= H^{(3)} \wedge \star F^{(3)}\\
    0 &= H^{(3)} \wedge F^{(5)}\\
    0 &= F^{(3)} \wedge F^{(5)}\\
    0 &= H^{(3)} \wedge F^{(3)}\\
    F^{(5)} &= \star F^{(5)}\\
    \Ric(X,Y) &= \tfrac12 F^{(1)}(X) F^{(1)}(Y) + \tfrac{1}{2} \inint{X}{Y}{H^{(3)}} + \tfrac12 \inint{X}{Y}{F^{(3)}}\\
      & \qquad {} + \tfrac14 \inint{X}{Y}{F^{(5)}} - \tfrac{1}{8}g(X,Y)
      \nrmsq{H^{(3)}} - \tfrac{1}{8} g(X,Y) \nrmsq{F^{(3)}}~,
  \end{aligned}
\end{equation}
where again for the case of a strongly symmetric background we have that in addition $F^{(1)}=0$.

The aim of this paper is to determine which ten-dimensional lorentzian symmetric spaces can carry invariant forms $F^{(1)}$, $F^{(3)}$, $F^{(5)}$ and $H^{(3)}$ satisfying equations \eqref{eq:IIBEoMsym} and to determine where possible the full moduli space of such backgrounds.  As mentioned above, strongly symmetric backgrounds are preserved by $\SL(2,\RR)$ duality transformations.  For the sake of economy we will not list strongly symmetric backgrounds that are so related, but rather we will use the action of $\SL(2,\RR)$ to set $\tau=i$ and then use the action of the $\SO(2)$ subgroup stabilising $\tau=i$ to further simplify the solution.  The understanding is that every background where $\fGa=0$ is to be thought of as a representative of its $\SL(2,\RR)$ orbit.

\subsection{Ten-dimensional lorentzian symmetric spaces}
\label{sec:ten-dimens-lorentz}

Ten-dimensional lorentzian symmetric spaces are easily listed based on the classifications of indecomposable lorentzian symmetric spaces and of irreducible riemannian symmetric spaces.  The discussion is very similar to that of \cite{FigueroaO'Farrill:2011fj} which discussed the eleven-dimensional case, so we will be brief.

A lorentzian locally symmetric space $(M,g)$ is locally isometric to a product
\begin{equation}
  \label{eq:decomposition}
  M = M_0 \times M_1 \times \ldots \times M_k
\end{equation}
where $M_0$ is an indecomposable lorentzian symmetric space and $M_{i>0}$ are irreducible riemannian symmetric spaces.

The irreducible riemannian symmetric spaces were classified by Élie Cartan (see, e.g., \cite{Helgason}).  Each symmetric space is determined locally by a pair of Lie algebras $(\mathfrak{g},\mathfrak{h})$, where $\fh$ is the fixed point set of an involutive automorphism of $\fg$.  This means that $\fg = \fh \oplus \fm$, where $[\fh,\fm] \subset \fm$ and $[\fm,\fm] \subset \fh$.  The subspace $\fm$ is a model for the tangent space at the origin to the symmetric space and the linear isotropy representation of $\fh$ on $\fm$ coincides the holonomy representation of the Levi-Civita connection.  It follows that the space of parallel forms in the symmetric space corresponding to $(\fg,\fh)$ is isomorphic to the $\fh$-invariant subspace of $\Lambda^{\ast} \fm$, as explained, e.g., in \cite{FigueroaO'Farrill:2011fj} in a similar context.

Table~\ref{tab:rsym} lists all the irreducible riemannian symmetric spaces of dimension less than or equal to $9$, together with the ranks of the parallel forms.  Each row in the diagram corresponds to two symmetric spaces: one compact and one noncompact.  The names in the last column correspond to the compact spaces and use the following notation $G^+_\RR(p,n)$ denotes the grassmannian of oriented real $p$-planes in $\RR^n$, $G_\CC(p,n)$ is the grassmannian of complex $p$-planes in $\CC^n$, $\ASSOC$ is the grassmannian of associative $3$-planes in $\RR^7$ and $\SLAG_n$ in the grassmannian of special lagrangian planes in $\CC^n$.  We also have to consider the one-dimensional symmetric space with metric $dt^2$ as a possible ingredient in the construction of general riemannian symmetric spaces.

\begin{table}[h!]
  \centering
  \renewcommand{\arraystretch}{1.3}
  \caption{Irreducible $d$-dimensional riemannian symmetric spaces with $d\leq 9$.  The names are those of the compact forms.}\label{tab:rsym}
  \begin{tabular}{>{$}r<{$}|>{$}c<{$}|>{$}c<{$}|>{$}c<{$}|>{$}l<{$}|>{$}l<{$}}
    \multicolumn{1}{c|}{dim} & \fg~\text{(compact)} & \fg~\text{(noncompact)} & \fh & \multicolumn{1}{c|}{$\fh$-inv. forms} & \multicolumn{1}{c}{Name}\\
    \hline
    2 & \fu(2) & \fu(1,1) & \fu(1) \oplus \fu(1) & 0,2 & S^2\\[3pt]
    3 & \fsu(2) \oplus \fsu(2) & \fsl(2,\CC) & \fsu(2) & 0,3 & S^3\\[3pt]
    4 & \fu(3) & \fu(2,1) & \fu(2) \oplus \fu(1) & 0,2,4 & \CP^2\\
    4 & \fsp(2) & \fsp(1,1) & \fsp(1) \oplus \fsp(1) & 0,4 & S^4\\[3pt]
    5 & \fsu(3) & \fsl(3,\RR) & \fso(3) & 0,5 & \SLAG_3\\ 
    5 & \fsu(4) & \fsl(2,\HH) & \fsp(2) & 0,5 & S^5\\[3pt]
    6 & \fu(4) & \fu(3,1) & \fu(3) \oplus \fu(1) & 0,2,4,6 & \CP^3\\
    6 & \fsp(2) & \fsp(2,\RR) & \fu(2) & 0,2,4,6 & G^+_\RR(2,5)\\
    6 & \fso(7)  & \fso(6,1)  & \fso(6) & 0,6 & S^6\\[3pt] 
    7 & \fso(8)  & \fso(7,1)  & \fso(7) & 0,7 & S^7\\[3pt]
    8 & \fu(4) & \fu(2,2) & \fu(2) \oplus \fu(2) & 0,2,4^2,6,8 & G_\CC(2,4)\\
    8 & \fu(5) & \fu(4,1) & \fu(4) \oplus \fu(1) & 0,2,4,6,8 & \CP^4 \\
    8 & \fso(9)  & \fso(8,1)  & \fso(8) & 0,8 & S^8\\
    8 & \fsp(3) & \fsp(2,1) & \fsp(2) \oplus \fsp(1) & 0,4,8 & \HP^2\\
    8 & \fg_{2(-14)} & \fg_{2(2)} & \fsp(1) \oplus \fsp(1) & 0,4,8 & \ASSOC \\
    8 & \fsu(3) \oplus \fsu(3) & \fsl(3,\CC) & \fsu(3) & 0,3,5,8 & \SU(3)\\[3pt]
    9 & \fsu(4) & \fsl(4,\RR) & \fso(4) & 0,4,5,9 & \SLAG_4\\
    9 & \fso(10) & \fso(9,1) & \fso(9) & 0,9 & S^9\\[3pt]
  \end{tabular}
\end{table}

The indecomposable lorentzian symmetric spaces are also classified \cite{CahenWallach,CahenParker} and are listed in Table~\ref{tab:lsym}.  They include, apart from the one-dimensional lorentzian factor with metric $-dt^2$, de~Sitter $dS_d$ and anti-de~Sitter $AdS_d$ spaces for $d>1$ and also the Cahen--Wallach spaces $\CW_d(\lambda)$ for $d>2$.  For a definition of $\CW_d(\lambda)$ and $\fg(\lambda)$ and $\fh(\lambda)$ see \cite[§2.2]{FigueroaO'Farrill:2011fj}.

\begin{table}[h!]
  \centering
  \renewcommand{\arraystretch}{1.3}
  \caption{Indecomposable $d$-dimensional lorentzian symmetric spaces.}
  \label{tab:lsym}
  \begin{tabular}{>{$}c<{$}|>{$}c<{$}|>{$}c<{$}|>{$}l<{$}}
    \text{type} & \fg & \fh & \multicolumn{1}{c}{$\fh$-invariant forms}\\
    \hline
    \dS_d & \fso(d,1) & \fso(d{-}1,1) & 1+t^d\\
    \AdS_d & \fso(d{-}1,2) & \fso(d{-}1,1) & 1+t^d\\
    \CW_d(\lambda) & \fg(\lambda) & \fh(\lambda) & 1+t(1+t)^{d-2} + t^d
  \end{tabular}
\end{table}

\subsection{Statistics of ten-dimensional lorentzian symmetric spaces}
\label{sec:stat-symm-backgr}

Let us count the number of (families of) eleven-dimensional lorentzian symmetric spaces.  Every indecomposable symmetric space (except for the Cahen--Wallach spaces) has a parameter corresponding to rescaling the metric.  As discussed above, indecomposable $d$-dimensional Cahen--Wallach spaces come in ($d-3$)-parameter families.  We will ignore these parameters in the counting, whence we will count families of geometries and not the geometries themselves.  Each such geometry is of the form $L_d \times R_{10-d}$, where $L$ is a $d$-dimensional indecomposable lorentzian symmetric space and $R$ is an ($10-d$)-dimensional riemannian symmetric space which is made out of the ingredients in Table~\ref{tab:rsym}.  Let $i_d$ denote the number of irreducible $d$-dimensional riemannian symmetric spaces up to local isometry.  Clearly $i_1 = 1$ since $\RR$ and $S^1$ are locally isometric.  The other values of $i_d$ can be read from Table~\ref{tab:rsym} and are tabulated in Table~\ref{tab:nirss}.

\begin{table}[h!]
  \centering
  \caption{Number of irreducible riemannian symmetric spaces up to local isometry}
  \label{tab:nirss}
  \renewcommand{\arraystretch}{1.3}
  \begin{tabular}{c|cccccccccc}
   $d$ & 1 & 2 & 3 & 4 & 5 & 6 & 7 & 8 & 9 \\
    \hline
    $i_d$ & 1 & 2 & 2 & 4 & 4 & 6 & 2 & 12 & 4
 \end{tabular}
\end{table}

We now let $r_d$ denote the number of $d$-dimensional riemannian symmetric spaces up to local isometry.  Clearly,
\begin{equation}
  \prod_{d=1}^\infty \frac{1}{1 - i_d t^d} = \sum_{d=1}^\infty r_d t^d~.
\end{equation}
Since we are interested only in $d\leq 9$, we can simply compute the first 9 terms in the left-hand side
\begin{equation}
  \prod_{d=1}^{9} \frac{1}{1 - i_d t^d} = 1 + t + 3 t^2 + 5 t^3 + 13 t^4 + 21 t^5 + 47 t^6 + 73 t^7 + 161 t^8 + 253 t^9 +  O\left(t^{10}\right)~,
\end{equation}
from where we can read off the values of $r_d$.  These are tabulated in Table~\ref{tab:nrss}.

\begin{table}[h!]
  \centering
  \renewcommand{\arraystretch}{1.3}
  \caption{Number of riemannian symmetric spaces up to local isometry}
  \label{tab:nrss}
  \begin{tabular}{c|cccccccccc}
    $d$ & 1 & 2 & 3 & 4 & 5 & 6 & 7 & 8 & 9\\
    \hline
    $r_d$ & 1 & 3 & 5 & 13 & 21 & 47 & 73 & 161 & 253
  \end{tabular}
\end{table}

Finally we let $\ell_d$ denote the number of indecomposable lorentzian symmetric spaces up to local isometry.  The total number of possible geometries is then
\begin{equation}
  N = \sum_{d=1}^{10} \ell_d r_{10-d}~.
\end{equation}
We notice that $\ell_1 = 1$, $\ell_2 = 2$ and $\ell_{d>2} = 3$, whence
\begin{equation}
  \begin{aligned}
    N &= 3 ( 1 + r_1 + r_2 + \cdots + r_7 ) + 2 r_8 + r_9\\
    &= 3 ( 1 + 1 + 3 + 5 + 13 + 21 + 47 + 73 ) + 2 \times 161 + 253 \\
    &= \boldsymbol{1067}~.
  \end{aligned}
\end{equation}

We do not need to consider separately the compact and noncompact forms of the riemannian factors, since this is determined from the sign of the curvature and this is in turn determined from the values of the fluxes.  This means that we actually have a total of 319 cases to consider.

\subsection{Methodology}
\label{sec:methodology}

First of all, we will further narrow down the list of possible geometries by first showing that there are no de~Sitter backgrounds and quickly listing those backgrounds having Cahen--Wallach or Minkowski factors.  This will leave about one hundred anti-de~Sitter geometries to consider, some of which can be easily ruled out.  This is the subject of the next section.

The remaining 60 geometries are studied in Section \ref{sec:analys-rema-spac}.  Each such geometry is characterised by a Lie algebra $\fg$ and an involutive automorphism with eigenspace decomposition $\fg = \fh \oplus \fm$.  The invariant metrics are the lorentzian inner products in $\fm$ which are invariant under the action of $\fh$.  This forms a cone inside $(S^2\fm)^\fh$.  We will parametrise those as well as the possible $F^{(1)} \in \fm^\fh$, $F^{(3)},H^{(3)} \in \left(\Lambda^3 \fm\right)^\fh$ and $F^{(5)} \in \left(\Lambda^5_+ \fm\right)^\fh$.  The field equations \eqref{eq:IIBEoMsym} will then become algebraic equations in those parameters which we will solve in most cases.

In practice, the metric parameters are fixed from the Einstein equations restricted to each of the irreducible factors in the decomposition \eqref{eq:decomposition}.  Indeed, for an irreducible symmetric space with symmetric pair $(\fg,\fh)$, the linear isotropy representation of $\fh$ on its complement $\fm$ is irreducible, hence any two invariant symmetric bilinear forms are proportional.  This means that the Ricci tensor, being an invariant symmetric bilinear form, must be proportional to the metric.  The proportionality constant, which is the associated metric parameter, is determined in terms of the flux parameters from the supergravity Einstein equation.  The metric on any flat factor can always be brought to standard form $(d\vartheta^1)^2 + (d\vartheta^2)^2 + \cdots$, whose resulting orthogonal symmetry can be used to further simplify the parametrisation of the forms.

\subsection{Relationship with symmetric M-theory backgrounds}
\label{sec:relat-with-symm}

Strongly symmetric type IIB backgrounds $(M,g)$ for which $\fGa=\fG=\fF=0$ (and $\tau=i$) are also strongly symmetric backgrounds of type IIA supergravity and hence can be oxidised to symmetric backgrounds of eleven-dimensional supergravity of the form $(M\times T^1, g + d\vartheta^2)$ and $F = \fH \wedge d\vartheta$.  This allows in some cases to compare with the results of \cite{FigueroaO'Farrill:2011fj}.  This will be highlighted when appropriate, particularly in Section \ref{sec:analys-rema-spac}.

\subsection{Curvature balance}
\label{sec:curvature-balance}

A useful heuristic, as explained in the context of eleven-dimensional supergravity in \cite[§4.1]{FigueroaO'Farrill:2011fj}, is that the Einstein equation for AdS backgrounds imposes a certain balance between the scalar curvatures of the AdS and riemannian factors.  In particular, at least one of the riemannian irreducible factors must have positive scalar curvature.  Let us prove this.  We will consider a background with geometry $AdS_d \times K^{10-d}$, where $K$ is a riemannian symmetric space, not necessarily irreducible.  The fields are given generically by
\begin{equation}
  \begin{aligned}[m]
    \fGa &= \alpha\\
    \fG &= \nu \wedge \beta^{(3-d)} + \gamma\\
    \fH &= \nu \wedge \delta^{(3-d)} + \epsilon\\
    \fF_+ &= \nu \wedge \theta^{(5-d)} + \star_K \theta~,
  \end{aligned}
\end{equation}
where $\alpha \in \Omega_{\text{inv}}^1(K)$, $\beta,\delta \in \Omega_{\text{inv}}^{3-d}(K)$, $\gamma,\epsilon \in \Omega_{\text{inv}}^3(K)$ and $\theta \in \Omega_{\text{inv}}^{5-d}(K)$, with the understanding that if $d>3$ then $\beta,\delta$ vanish and if $d>5$ then $\theta$ vanishes.  The Einstein equation along $K$ says that
\begin{multline}
  R_{ab} = \half \alpha_a \alpha_b - \half \left<\beta_a,\beta_b\right> + \half \left<\gamma_a,\gamma_b\right> - \half \left<\delta_a,\delta_b\right> + \half \left<\epsilon_a,\epsilon_b\right> - \tfrac14 \left<\theta_a,\theta_b\right>\\
  + \tfrac14 \left<(\star\theta)_a,(\star\theta)_b\right> - \tfrac18 g_{ab} (-|\beta|^2 + |\gamma|^2 - |\delta|^2 + |\epsilon|^2)~,
\end{multline}
in a notation where $\omega_a$ means the contraction of a form $\omega$ on $K$ with the $a^{\mathrm{th}}$ element of the frame.  Tracing with $g^{ab}$ we see that the scalar curvature $\mathrm{scal}_K$ of $K$ is given by
\begin{equation}
  \mathrm{scal}_K = \half|\alpha|^2 + \frac{d+2}8 (|\gamma|^2 + |\epsilon|^2) + \frac{d}4 |\theta|^2 + \frac{3d-2}8 (|\beta|^2 + |\delta|^2)~,
\end{equation}
which is manifestly non-negative.  The only way this can vanish (given that $d>1$) is for all the fields to vanish, whence the scalar curvature for $AdS_d$ would also vanish, which is absurd.  Therefore $\mathrm{scal}_K > 0$.

\subsection{Further observations}
\label{sec:further-observations}

We collect two further useful observations.  The first is that as far as the search for backgrounds is concerned, namely as far as all we are interested in is to find a solution of the supergravity field equations, the spaces in each of the pairs $(S^5, SLAG_3)$, $(\CP^3, G_\RR^+(2,5))$ and $(\HP^2, ASSOC)$ are interchangeable, since they have the same invariant forms which can moreover be identically normalised.

A second observation, related to the previous one, is that the existence of a background with an $S^d$ irreducible factor implies the existence of a background where $S^d$ is replaced by any other (not necessarily irreducible) riemannian symmetric space of the same dimension and the same scalar curvature.  For example, any background with an $S^5$ factor implies the existence of a background where the $S^5$ has been replaced by $S^3 \times S^2$, provided that the Einstein equations for the $S^3$ and $S^2$ factors have the same constant as that of the $S^5$ Einstein equation, and similarly any $S^4$ factor in a background can be replaced by $S^2 \times S^2$, provided that that curvatures are the same as for the $S^4$ and in this way obtain a new background.  A similar observation applies to the $AdS_d$ factor in a background, which can be replaced by a lower-dimensional $AdS_p$ and some noncompact riemannian symmetric space of dimension $d-p$ and still obtain a background.  For example, any $AdS_5$ background implies the existence of a background where the $AdS_5$ is replaced with $AdS_3 \times H^2$ or $AdS_2 \times H^3$, provided that the Einstein equations of all spaces are the same.  Similarly, an $AdS_4$ background implies the existence of a background where $AdS_4$ is replaced by suitably curved $AdS_2 \times H^2$.  The reason in all cases is the same: all invariant forms on $S^d$ or $AdS_d$ are multiples of the volume form, which can be substituted for the volumes form of $S^p \times S^{d-p}$ or $AdS_p \times H^{d-p}$, et cetera.  More generally, in any AdS background, any factor $M_i$ can be substituted by any symmetric space (not necessarily irreducible) which admits invariant forms of the same ranks and normalisations as those of $M_i$.  In practice and due to the low dimension, this only applies to the examples mentioned above.

\subsection{Notation}
\label{sec:notation}

Unless explicited stated otherwise, the following notation is used in this paper.  Volume forms will be denoted $\nu$ and they come adorned with a subscript consistent with the local decomposition of $M$ given in equation \eqref{eq:decomposition}.  In other words, $\nu_0$ will denote the volume form of the indecomposable lorentzian factor, $\nu_1$ that of the first irreducible riemannian factor, et cetera.  The only exception to this rule is that whenever there are flat riemannian factors in the decomposition, we will consider them to be an ``irreducible'' factor and write them last.  So for example, if $M = AdS_3 \times S^3 \times T^4$ we will let $\nu_0$, $\nu_1$ and $\nu_2$ denote the volume forms on $\AdS_3$, $S^3$ and $T^4$, respectively.  We will let $g_0$, $g_1$,... similarly denote the metrics of $M_0$, $M_1$,... whose Ricci curvatures are denoted $R_0$, $R_1$,... respectively.  In this way the $M_0$ factor of the Einstein equation will read $R_0 = \lambda g_0$, for some $\lambda$.  The flat coordinates on the flat riemannian directions will be denoted by $\vartheta^a$.  Finally, other notation will be introduced as needed.

\section{Analysis of special cases}
\label{sec:analys-spec-cases}

We will first deal with those geometries where $M_0$ in equation \eqref{eq:decomposition} is either de~Sitter space, one-dimensional or a Cahen--Wallach space.  We will then rule out geometries where either $M_0$ or one of the $M_i$ in equation \eqref{eq:decomposition} is of dimension $d>6$.

\subsection{No de~Sitter backgrounds}
\label{sec:de-sitter}

We look at backgrounds of the form $dS_d \times M^{10-d}$, for $d\geq 2$, and analyse the restriction to $dS_d$ of the Einstein equation.  To this end let us take Latin indices $a,b,\ldots$ to be over the de~Sitter part and Latin indices $i,j,\ldots$ to be over the riemannian part.  Additionally, a bar denotes the absence of a $dS_d$ factor.  The notation $\tau^{(r)}$ (perhaps further adorned) denotes a rank-$r$ invariant form on the riemannian factor.

The most general 5-form we can construct (only available for $d \leq 5$) takes the form
\begin{equation}
  \fF = \nu_0^{(d)} \wedge \tau_{5}^{(5-d)} + \overline{F}^{(5)}~,
\end{equation}
with $\nu_0^{(d)}$ the volume form on $dS_d$, yielding
\begin{equation}
  \inint{a}{b}{\fF} = \inint{a}{b}{\nu_0^{(d)}} \nrmsq{\tau_{5}^{(5-d)}} = -\nrmsq{\tau_{5}^{(5-d)}} g_{a b}~.
\end{equation}
Similarly, the most general 3-form we can construct (only available for $d \leq 3$) takes the form
\begin{equation}
  K^{(3)} = \nu_0^{(d)} \wedge \tau_{K}^{(3-d)} + \overline{K}^{(3)}~,
\end{equation}
where $K$ can stand for either $F$ or $H$, yielding
\begin{equation}
  \inint{a}{b}{K^{(3)}} = \inint{a}{b}{\nu_0^{(d)}} \nrmsq{\tau_{K}^{(3-d)}} = -\nrmsq{\tau_{K}^{(3-d)}} g_{a b}
\end{equation}
and
\begin{equation}
\nrmsq{K^{(3)}} = \nrmsq{\nu_0^{(d)}} \nrmsq{\tau_{K}^{(3-d)}} + \nrmsq{\overline{K}^{(3)}}~.
\end{equation}
Finally, as there can be no 1-forms with legs in the de~Sitter part,
\begin{equation}
	\fGa(X)\fGa(Y) = 0
\end{equation}
for any vector fields $X,Y$ tangent to $dS_d$.

In summary, the Einstein equation on the de~Sitter part says
\begin{equation}
  R_{a b} = -\tfrac{1}{8} \left( 3 \nrmsq{\tau_{H}^{(3-d)}} + 3 \nrmsq{\tau_{F}^{(3-d)}} + 2 \nrmsq{\tau_{5}^{(5-d)}} + \nrmsq{\overline{H}^{(3)}} + \nrmsq{\overline{F}^{(3)}} \right)g_{a b}~.
\end{equation}
But notice that the expression in parenthesis is positive-semidefinite, contradicting the fact that $dS_d$ has positive scalar curvature.  Hence this rules out all backgrounds with a de~Sitter part.  Since we only used the Einstein equation, this result holds more generally for homogeneous backgrounds, even if the underlying homogeneous geometry is not symmetric.

\subsection{Backgrounds with one-dimensional lorentzian factor}\hypertarget{geom:d1}{}
\label{sec:backgrounds-with-one}

The metric here takes the form $g = -dt^2 + \overline{g}$.  We will analyse the restriction of the Einstein equation to the $dt$ part of such a background.  A bar denotes the absence of a $dt$ factor.  Let $K$ stand for either of $\fH$,  $\fG$ or $\fF$ and let $K_t = \iota_{\frac{\partial}{\partial t}} K$.  Then $K$ may or may not have legs along $dt$, whence it can take one of the following two forms:
\begin{enumerate}
\item $K = \overline{K}$, whence $K_t = 0$ and $\nrmsq{K} = \nrmsq{\overline{K}}$; and
\item $K = dt \wedge \overline{K} + \overline{L}$, whence $K_t = \overline{K}$ and hence $\left<K_t,K_t\right> = \nrmsq{\overline{K}}$ and $\nrmsq{K} = \nrmsq{\overline{L}} - \nrmsq{\overline{K}}$.
\end{enumerate}
Similarly, $\fGa$ can take one of the following two forms:
\begin{enumerate}
\item $\fGa= \overline{K}$, whence $\fGa_t=0$; and
\item $\fGa = f\, dt + \overline{K}$, whence $\fGa_t = f$,
\end{enumerate}
where as before we have defined $\fGa_t = \iota_{\frac{\partial}{\partial t}} \fGa$.

Since $t$ is a flat direction, the $tt$ component of the Ricci curvature  vanishes, whence
\begin{equation}
0 \stackrel{!}{=} R_{tt} = \tfrac{1}{2}\fGa_t\fGa_t + \tfrac{1}{2} \left<\fH_t,\fH_t\right> + \tfrac{1}{2} \left<\fG_t,\fG_t\right> + \tfrac{1}{4} \left<\fF_t,\fF_t\right> + \tfrac{1}{8} \nrmsq{\fH} + \tfrac{1}{8} \nrmsq{\fG}~.
\end{equation}
It is clear both from the pairing of inner interior products and norms of the 3-forms in the Einstein equation that the right hand side will always be a sum of positive-coefficient norms of forms in the complement factor.  These norms are all positive semidefinite, and thus if we find that if a norm contributes to $R_{tt}$ then it must be zero.

We consider first the contributions from the 3-forms, where $K$ can be either $\fG$ or $\fH$:
\begin{enumerate}
	\item $K = \overline{K} \then \frac{1}{2} \left<K_t,K_t\right> + \frac{1}{8} \nrmsq{K} = \frac{1}{8} \nrmsq{\overline{K}} \then \nrmsq{\overline{K}} = 0 \then K = 0$
	\item $K = dt \wedge \overline{K} + \overline{L} \then \frac{1}{2} \left<K_t,K_t\right> + \frac{1}{8} \nrmsq{K} = \frac{3}{8} \nrmsq{\overline{K}} + \frac{1}{8} \nrmsq{\overline{L}} \then \nrmsq{\overline{K}} = \nrmsq{\overline{L}} = 0 \newline \then \overline{K} = \overline{L} = 0 \then K = 0$
\end{enumerate}
Thus the 3-forms must be zero.

Next we consider the contribution from the 5-form $\fF$:
\begin{enumerate}
	\item $\fF = \overline{K} \then \frac{1}{4} \left<\fF_t,\fF_t\right> = 0$
	\item $\fF = dt \wedge \overline{K} + \overline{L} \then \frac{1}{4} \left<\fF_t,\fF_t\right> = \frac{1}{4} \nrmsq{\overline{K}} \then \nrmsq{\overline{K}} = 0 \then \fF = \overline{L}$
\end{enumerate}
Thus the 5-form can have no legs along $dt$, which contradicts the self-duality.  Indeed, a self-dual 5-form has zero norm, hence if $\fF = \overline{K}$, then $0 = \nrmsq{\fF} =\nrmsq{\overline{K}}$, whence $\overline{K} = 0$.

Finally, we consider the contribution from the 1-form.  The $tt$ Einstein equation says that the $dt$ factor must vanish, whence $\fGa = \overline{K}$ for some one-form $\overline{K}$ on the riemannian part. But now the first equation in \eqref{eq:IIBEoMhom} says that $\nrmsq{\overline{K}}=0$, whence $\fGa = 0$.  Notice that this is true more generally for any homogeneous background and not just those where the underlying homogeneous space is symmetric.

In summary, we have seen that the only solution is $\fGa = \fH = \fG = \fF = 0$.  This means that $\overline{g}$ is Ricci-flat and hence flat.  This is true more generally than in the symmetric case, since any homogeneous Ricci-flat riemannian manifold is flat \cite{AlekseevskyKimelfeld}.  The only homogeneous solution is therefore locally isometric to the Minkowski vacuum.

\subsection{Backgrounds with Cahen--Wallach factors}\hypertarget{geom:cw}{}
\label{sec:cw}

We look at backgrounds of the form $CW_{d}(\lambda) \times M^{10-d}$ and analyse the restriction of the Einstein equation to the transverse directions of the $CW_{d}(\lambda)$ space and also to the riemannian part.  We will take Latin indices $a,b,\ldots$ to be the transverse indices inside $CW_{d}(\lambda)$.  Additionally, a bar denotes the absence of a $CW_{d}(\lambda)$ component.

The most general invariant rank-$r$ form in this space has the form
\begin{equation}
  \label{eq:CWgeneralform} 
  W^{(r)} = \nu_0^{(d)} \wedge \overline{W}_1^{(r-d)} + \overline{W}_2^{(r)} + \sum_i  \tensor{\tau}{_{i}^{(s_i)}} \wedge \tensor{\overline{Z}}{_{i}^{(r-s_i)}}~,
\end{equation}
where the first term is the wedge product of the Cahen--Wallach volume form with an invariant form on as riemannian factor, the second term is an invariant form on a riemannian factor, and the third term is a sum of wedge products of Cahen--Wallach non-volume invariant forms and invariant forms on riemannian factors.  Using that the $\tau_i$ are null, the norm squared of $W^{(r)}$ is
\begin{equation}
\nrmsq{W^{(r)}} = - \nrmsq{\overline{W}_1^{(r-d)}} + \nrmsq{\overline{W}_2^{(r)}}~,
\end{equation}
whereas the inner interior product in the transverse directions of the Cahen--Wallach part is
\begin{equation}
  \inint{a}{b}{W^{(r)}} = - \nrmsq{\overline{W}_1^{(r-d)}} g_{a b}~.
\end{equation}

Using that the Cahen--Wallach spaces are Ricci-null, so that $R_{ab} = 0$, those components of the Einstein equation yield
\begin{equation}
  3 \nrmsq{\overline{H}_1^{(3-d)}} + \nrmsq{\overline{H}_2^{(3)}} + 3 \nrmsq{\overline{F}_1^{(3-d)}} +  \nrmsq{\overline{F}_2^{(3)}} + 2 \nrmsq{\overline{F}_1^{(5-d)}} \stackrel{!}{=} 0~.
\end{equation}
This shows that $\nrmsq{\fH} = \nrmsq{\fG} = 0$, whence $\fH$ and $\fG$, if nonzero, are null.  The first equation in \eqref{eq:IIBEoMhom} then says that $\nrmsq{\fGa} = 0$, whence since $\overline{F}_1^{(5-d)}=0$, it says that $\overline{F}_2^{(1)} = 0$ as well.  Finally, since the above equation says that $\fF$ has no component with a Cahen--Wallach volume form, the fact that $\nrmsq{\fF} = 0$ (which follows from self-duality), implies that $\overline{F}^{(5)}_2 = 0$.  In summary, all of $\fH$, $\fGa$, $\fG$ and $\fF$ take the form \eqref{eq:CWgeneralform} where the first two terms are absent.  But because the $\tau_i$ are null, they do not contribute to the Ricci tensor of the riemannian factor, which is therefore forced to be Ricci-flat and hence flat.  We are thus left with only $CW_{d}(\lambda) \times \mathbb{R}^{10-d}$, which we can think of a (degenerate) Cahen--Wallach spacetime.

\subsection{High-dimensional riemannian factors}
\label{sec:high-dimens-riem}

We denote a rank $r$ invariant form in the riemannian factor by $\tau^{(r)}$ and an $AdS_d$ volume form by $\nu_0^{(d)}$.

Three of the field equations \eqref{eq:IIBEoMsym} involve only the two 3-forms $\fH$ and $\fG$.  We first note that if $\fH$ is zero, then it has zero norm and thus forces $\fGa = 0$ and $\fG$ to have zero norm because, having already dealt with the $d=1$ and Cahen--Wallach cases, $\fGa$ can only have legs along the riemannian factor.  Furthermore, if $\fH = \fG$, then $\fGa = \nrmsq{\fH} = \nrmsq{\fG} = 0$.

If we suppose that a space only admits a one-parameter family of invariant 3-forms, then these invariant 3-forms must have zero norm.  However, the only spaces with any invariant forms of zero norm are the Cahen--Wallach spaces.  We have already analysed these and thus we rule out any further space that has only a one-parameter family of invariant 3-forms (and no 5-forms).

\subsubsection*{9-dimensional riemannian factors:}

These are precisely the cases already treated in Section~\ref{sec:backgrounds-with-one}.

\subsubsection*{8-dimensional riemannian factors:}

\begin{itemize}\hypertarget{geom:AdS2SU3}{}
\item $\mathbf{( \mathfrak{su}(3) \oplus \mathfrak{su}(3), \mathfrak{su}(3) )}$	and $\mathbf{( \mathfrak{sl}(3,\mathbb{C}), \mathfrak{su}(3) )}$\\
  The only available complementary space is $AdS_2$.  The Lie group $SU(3)$ is rationally homotopy equivalent to $S^3 \times S^5$, whence it has an invariant $3$-form $\tau^{(3)}$ and an invariant $5$-form $\tau^{(5)} = \star \tau^{(3)}$.  As there are no invariant 1-forms, $\fGa=0$, whereas since the 3-form field equations cannot be satisfied with only one (riemannian) 3-form, also $\fH = \fG = 0$.  The most general self-dual 5-form is
  \begin{equation}
    \fF = \gamma(\tau^{(5)} - \tau^{(3)} \wedge \nu_0^{(2)})~.
  \end{equation}
  After a brief calculation, and with the normalisation $|\tau^{(3)}|^2=1$, the Einstein equation yields
  \begin{equation}
    R_0 = -\tfrac14 \gamma^2 g_0 \qquad\text{and}\qquad R_1 = \tfrac1{16} \gamma^2 g_1~,
  \end{equation}
  which gives a solution for the compact case $AdS_2 \times SU(3)$.
\item All other 8-dimensional riemannian factors:\\
  The only available complement space is $AdS_2$.  As such, there is no way to construct any 3-forms or 5-forms and Ricci-flatness is forced.  However, these spaces are not Ricci-flat and so we have a contradiction.  These spaces are thus ruled out.
\end{itemize}

\subsubsection*{7-dimensional riemannian factors:}

There are no invariant 3-forms or 5-forms in the riemannian part.  All invariant 3-forms are proportional to the volume form of the lorentzian factor, but a single 3-form cannot satisfy the field equations unless it vanishes.  But in that case the solution has no fluxes and the Einstein equation forces it to be Ricci-flat, contradicting the fact that no irreducible riemannian symmetric space of dimension greater than one is Ricci-flat.  Therefore there are no such backgrounds.

\subsection{High-dimensional \texorpdfstring{$AdS_d$}{AdS} factors}
\label{sec:high-dimens-ads}

\subsubsection*{The case \texorpdfstring{$d>7$}{d>7}:}

In this case, there is no way to construct any nonzero 3-forms or 5-forms and the Einstein equations force Ricci-flatness, again contradicting the negative curvature of $AdS_d$.

\subsubsection*{The case \texorpdfstring{$d=7$}{d=7}:}

In this case, the only invariant 3-forms are proportional to the volume form of the complementary riemannian factor.  This means that the field equations are only satisfied if they vanished, but then the Einstein equation would force Ricci-flatness in contradiction with the negative curvature of $AdS_7$.

\subsection{Summary of geometries ruled out thus far}
\label{sec:summ-geom-ruled}

Here we list the geometries that we have ruled out in this section.  In those cases where we list a compact riemannian symmetric space, the noncompact dual is ruled out as well.  The notation is such that $X$ is any lorentzian symmetric space and $Y$ is any riemannian symmetric space and we use $\RR^{0,1}$ for a one-dimensional lorentzian manifold with metric $-dt^2$ and $\RR^d$ for a $d$-dimensional flat riemannian manifold.
\begin{multicols}{2}
\begin{itemize}
\item $dS_d \times Y^{10-d}$
\item $\RR^{0,1} \times Y^{9}$ with $Y \neq \mathbb{R}^9$
\item $CW_{d}(\lambda) \times Y^{10-d}$ with $Y^{10-d} \neq \mathbb{R}^{10-d}$
\item $S^7 \times X^{3}$
\item $G_{\mathbb{C}}(2,4) \times X^{2}$
\item $\mathbb{CP}^4 \times X^{2}$
\item $S^8 \times X^{2}$
\item $\mathbb{HP}^2 \times X^{2}$
\item $\ASSOC \times X^{2}$
\item $SU(3) \times X^{2}$ with $X^{2} \neq AdS_2$
\item $AdS_{d\geq7} \times Y^{10-d}$
\end{itemize}
\end{multicols}

\section{Remaining AdS backgrounds}
\label{sec:analys-rema-spac}

We are left with a set of spaces of the form $AdS_{d<7} \times M^{10-d}$ where all irreducible riemannian factors in $M$ have dimension $k<7$.  As explained earlier, we need not distinguish between the compact and noncompact form of any irreducible riemannian factor as this can be deduced later from the restriction to that factor of the Einstein equation.  Restricting further to spaces that have either non-zero invariant self-dual 5-forms or at least a two-dimensional space of invariant 3-forms, we are left with 60 spaces.

\begin{table}[H]
\tabulinesep=3pt
\begin{tabu}{X[l] *5{X[c]}} \firsthline	
  $\mathbf{AdS_d}$ & $AdS_2$ & $AdS_3$ & $AdS_4$ & $AdS_5$ & $AdS_6$\\
\textbf{\# spaces} & 23	& 19 & 9 & 8 & 1\\\lasthline
\end{tabu}
\end{table}

\subsection{Polynomial systems}
\label{sec:polynomial-systems}

The field equations are reduced with parametrised sums of invariant forms resulting in a system of polynomial equations in said parameters.  We then add constraints on the system from geometrical considerations, such as requiring that Ricci-flat and non-Ricci-flat factor geometries have respectively vanishing and non-vanishing Ricci-tensor restrictions.  Solving this final system gives us the moduli space of the background.

In general, a given candidate geometry will have spaces of invariant 1-forms, 3-forms and self-dual 5-forms of dimensions $m_1 = \dim \fm^\fh$, $m_3 = \dim \left(\Lambda^3\fm\right)^\fh$ and $m_5^+= \dim \left(\Lambda_+^5\fm\right)^\fh$, respectively.  This gives us a total of $m_1 + 2 m_3 + m_5^+$ parameters which are then constrained by the field equations to form the moduli space.

\subsubsection*{Analytical solutions:}
It is desirable to solve for the moduli space exactly.  In many cases this can be done, and we simply solve symbolically over the reals.  We denote by $\alpha_i$, $\beta_j$, $\gamma_k$ and $\kappa_\ell$ the parameters in $\fH$, $\fG$, $\fF$ and $\fGa$, respectively.  We denote by $R_n$ and $g_n$, respectively, the restrictions of the Ricci tensor and metric tensor to $M_n$ in the decomposition \eqref{eq:decomposition}.

\subsubsection*{Numerical solutions:}  When our system becomes unwieldy, analytical solution is no longer an option (even after using the homothety invariance described in equation \eqref{eq:homothety}) and we must sadly search for numerical solutions \cite{MR2499505}.

Our technique is blunt: we take the sum of the squares of our normalised polynomial system $F=\sum_{i} f_i^2$ and then use a low discrepancy quasi-random sampling of the homothetically compactified solution space of our system as seeds for standard numerical minimisation routines applied to $F$.  We accept local minima as valid solutions as long as $\lvert F \rvert < 10^{-30}$.  Note that calculations were carried out with a working precision of $10^{-60}$.  Checks using the application of this technique to the polynomial systems that were analytically solvable were encouraging.  However, a pinch of salt is prescribed.

We applied this technique in two ways to help us with difficult polynomial systems.  First, to trawl the solution space of a system to hint at whether solutions may exist and if so, to indicate the (non-)compactness of factor geometries.  Second, and when solutions are suggested to exist, to present potential ans\"atze for finding exact solutions.

\subsubsection*{Limit solutions:}
If we have an exactly solved moduli space for a particular background, we may see that in certain limits we get another background, generally with part of the original geometry becoming flat.  In this way we can see that certain backgrounds must exist even if we are unable to determine fully their moduli space.  In these cases, we may not even look for numerical solutions because there is nothing further to gain.  In particular, by considering the balance of curvatures between factors, we know that we do not miss any non-compact factor geometries by doing this.

We will only list those geometries for which we have found solutions.  Those inadmissible geometries which we have managed to rule out by explicit calculation and not merely by general arguments are listed in Appendix \ref{sec:inadm-geom}.

\subsection{\texorpdfstring{$\mathbf{AdS_5}$}{AdS d=5} backgrounds}

\subsubsection*{\texorpdfstring{$\mathbf{AdS_5 \times S^5}$ {\bf and} $\mathbf{AdS_5 \times SLAG_3}$}{AdS5 x S5 and AdS5 x SLAG3}}\hypertarget{geom:AdS5S5}{}
The field equations admit the following solution:
\begin{equation}
  \begin{aligned}[m]
    \fGa &= \fG = \fH = 0\\
    \fF &= \gamma(\nu_0 - \nu_1)~,
  \end{aligned}
\end{equation}
in the notation introduced in Section \ref{sec:notation}.  The Einstein equation yields
\begin{equation}
  R_0 = -\tfrac14 \gamma^2 g_0\qquad\text{and}\qquad R_1 = \tfrac14 \gamma^2 g_1~,
\end{equation}
giving a solution for the compact cases $AdS_5 \times S^5$ and $AdS_5 \times SLAG_3$.

\subsubsection*{\texorpdfstring{$\mathbf{AdS_5 \times S^3 \times S^2}$}{AdS5 x S3 x S2}}\hypertarget{geom:AdS5S3S2}{}

The field equations admit the following solution:
\begin{equation}
  \begin{aligned}[m]
    \fGa &= \fG = \fH =  0\\
    \fF &= \gamma(\nu_0 - \nu_1 \wedge \nu_2)~.
  \end{aligned}
\end{equation}
The Einstein equation yields
\begin{equation}
  R_0 = -\tfrac14 \gamma^2 g_0\qquad R_1 = \tfrac14 \gamma^2 g_1\qquad\text{and}\qquad R_2 = \tfrac14 \gamma^2 g_2~,
\end{equation}
giving a solution for the compact case $AdS_5 \times S^3 \times S^2$.  The existence of such a background follows from the existence of the $AdS_5 \times S^5$ background by the second observation in Section~\ref{sec:further-observations}.

\subsection{\texorpdfstring{$\mathbf{AdS_4}$}{AdS d=4} backgrounds}

\subsubsection*{\texorpdfstring{$\mathbf{AdS_4 \times S^3 \times S^2 \times T^1}$}{AdS4 x S3 x S2 x T1}}\hypertarget{geom:AdS4S3S2T1}{}

The field equations admit the following solution, with $\xi_{1,2} = \pm1$:
\begin{equation}
  \begin{aligned}[m]
    \fGa &= \kappa d\vartheta\\
    \fG &= 0\\
    \fH &= \xi_1 \sqrt2 \kappa \nu_2 \wedge d\vartheta\\
    \fF &= \xi_2 \sqrt5 \kappa (\nu_0 \wedge d\vartheta + \nu_1 \wedge \nu_2)
  \end{aligned}
\end{equation}
with $\vartheta$ a local coordinate on $T^1$ and where the Einstein equation yields
\begin{equation}
  R_0 = -\tfrac{3}{2}\kappa^2 g_0\qquad R_1 = \kappa^2 g_1 \qquad\text{and}\qquad R_2 = 2\kappa^2 g_2~,
\end{equation}
giving a solution for $AdS_4 \times S^3 \times S^2 \times T^1$.  This example shows that the second observation in Section~\ref{sec:further-observations} cannot be used in reverse; namely we cannot deduce the existence of an $AdS_4 \times S^5 \times T^1$ background, since the above background uses not just the volume form on $S^3 \times S^2$, but in fact also the volume 2-form on $S^2$ which is not available in $S^5$.  Nevertheless, this solution \emph{can} be used to obtain a symmetric background with underlying geometry $AdS_2 \times H^2 \times S^3 \times S^2 \times T^1$.

\subsection{\texorpdfstring{$\mathbf{AdS_3}$}{AdS d=3} backgrounds}

For $AdS_3$ and $AdS_2$ the moduli spaces become increasingly difficult to compute as the number of 3-form components increases, and even when computed may be difficult to interpret.

\subsubsection*{\texorpdfstring{$\mathbf{AdS_3 \times S^5 \times S^2}$ \bf{and} $\mathbf{AdS_3 \times SLAG_3 \times S^2}$}{AdS3 x S5 x S2 and AdS3 x SLAG3 x S2}}\hypertarget{geom:AdS3S5H2}{}

The field equations admit the following solution:
\begin{equation}
  \begin{aligned}[m]
    \fGa &= \fG = \fH = 0\\
    \fF &= \gamma(\nu_1 - \nu_0 \wedge \nu_2)~.
  \end{aligned}
\end{equation}
The Einstein equation yields:
\begin{equation}
  R_0 = -\tfrac14 \gamma^2 g_0\qquad R_1 = \tfrac14 \gamma^2 g_1 \qquad\text{and}\qquad R_2 = -\tfrac14 \gamma^2 g_2~,
\end{equation}
giving a solution for the cases $AdS_3 \times S^5 \times H^2$ and $AdS_3 \times SLAG_3 \times H^2$.  We could have inferred the existence of such a background from that of the $AdS_5 \times S^5$ background by the second observation in Section~\ref{sec:further-observations}.

\subsubsection*{\texorpdfstring{$\mathbf{AdS_3 \times S^4 \times S^3}$}{AdS3 x S4 x S3}}\hypertarget{geom:AdS3S4S3}{}

The field equations admit the following solutions with $\xi = \pm 1$:
\begin{equation}
  \begin{aligned}[m]
    \fGa &= \fF = 0\\
    \fG &= \beta(\nu_0 + \xi \nu_2)\\
    \fH &= \alpha(\nu_0 + \xi \nu_2)~.
  \end{aligned}
\end{equation}
As explained in the introduction, there is a residual $SO(2)$ subgroup of the $\SL(2,\RR)$ duality group which we may use to simplify the solution further.  This subgroup acts by rotations in the $(\fG,\fH)$ plane or, equivalently, in the $(\alpha,\beta)$ plane.  Hence we may and will use this to set $\beta=0$, whence the solutions are
\begin{equation}
  \label{eq:fga-=-fg}
  \fGa = \fG = \fF = 0 \qquad\text{and}\qquad \fH = \alpha(\nu_0 \pm \nu_2)~.
\end{equation}
The Einstein equation yields:
\begin{equation}
  R_0 = -\tfrac{1}{2} \alpha^2 g_0\qquad R_1 = 0\qquad R_2 = \tfrac{1}{2} \alpha^2 g_2~.
\end{equation}
As $R_1 = 0$ is forced, this geometry is ruled out and what we have is a background with underlying geometry $AdS_3 \times T^4 \times S^3$.

To be sure, what we have done is write down one representative background in its   $\SL(2,\RR)$ orbit.  We obtain the other backgrounds in the same duality orbit by applying a general duality transformation. It may be instructive to show how this works in this one example.  The $\SL(2,\RR)$ duality transformations are described in Section \ref{sec:sl2R}.  In more detail, under a general element
\begin{equation}
  \begin{pmatrix} a & b\\ c & d \end{pmatrix} \in SL(2,\mathbb{R})~,
\end{equation}
and \emph{in a strongly symmetric background} with $\tau=i$ and $\fG=0$ such as this one, the transformed background has both the metric and $\fF$ unchanged, whereas the rest of the fields are
\begin{equation}
  \begin{aligned}
    (C^{(0)})' &= \frac{ac + bd}{c^2 + d^2}\\
    \phi' &= \ln (c^2 + d^2)\\
    (\fH)' &= a \fH\\
    (\fG)' &= (c^3 + c(d^2 -a^2) - abd) \fH~,
  \end{aligned}
\end{equation}
where $\fH$ is given in equation \eqref{eq:fga-=-fg}.

\subsubsection*{\texorpdfstring{$\mathbf{AdS_3 \times \mathbb{CP}^2 \times S^3}$}{AdS3 x CP2 x S3}}\hypertarget{geom:AdS3CP2S3}{}

The field equations admit the following solution with $\xi$ a sign, and where we have used the residual $\SO(2)$ duality to set $\fG=0$:
\begin{equation}
  \begin{aligned}[m]
    \fGa &= \fG = 0\\
    \fH &= \alpha(\nu_0 + \xi \nu_2)\\
    \fF &= \half (1 + \xi)\gamma(\nu_0 + \xi \nu_2) \wedge \omega_{\mathbb{CP}^2}
  \end{aligned}
\end{equation}
The Einstein equation then yields in both cases:
\begin{equation}
  R_0 = -\tfrac{1}{2} (\alpha^2 + \gamma^2) g_0\qquad R_1 = 0\qquad R_2 = \tfrac{1}{2} (\alpha^2 + \gamma^2) g_2~.
\end{equation}
As $R_1 = 0$ is forced, this geometry is ruled out and what we get is a background with underlying geometry $AdS_3 \times T^4 \times S^3$.

\subsubsection*{\texorpdfstring{$\mathbf{AdS_3 \times S^3 \times S^2 \times S^2}$}{AdS3 x S3 x S2 x S2}}\hypertarget{geom:AdS3S3S2H2}{}

The field equations first admit the following solution:
\begin{equation}
  \begin{aligned}[m]
    \fGa &= \fG = \fH = 0\\
    \fF &= \gamma_1 (\nu_0 \wedge \nu_2 - \nu_1 \wedge \nu_3) + \gamma_2 (\nu_0 \wedge \nu_3 - \nu_1 \wedge \nu_2)~.
  \end{aligned}
\end{equation}
The Einstein equation then yields:
\begin{equation}
  R_0 = -\tfrac14 (\gamma_1^2+\gamma_2^2) g_0\qquad R_1 = \tfrac14 (\gamma_1^2+\gamma_2^2) g_1\qquad R_2 = -\tfrac14 (\gamma_1^2-\gamma_2^2) g_2\qquad R_3 = \tfrac14 (\gamma_1^2-\gamma_2^2) g_3~,
\end{equation}
which gives a solution for $AdS_3 \times S^3 \times S^2 \times H^2$.  This solution degenerates to a solution for $AdS_3 \times S^3 \times T^4$ when $\gamma_1^2 = \gamma_2^2$.

The existence of the special case $\gamma_1=0$ of this background follows from the $AdS_5 \times S^5$ background by the second observation in Section~\ref{sec:further-observations}.

The field equations also admit the following solution with $\xi = \pm 1$:
\begin{equation}
  \begin{aligned}[m]
    \fGa &= \fG = 0\\
    \fH &= \alpha (\nu_0 + \xi \nu_1)\\
    \fF &= \gamma (\nu_0 - \nu_1) \wedge (\nu_2 - \xi \nu_3)~.
  \end{aligned}
\end{equation}
The Einstein equation then yields:
\begin{equation}
  R_0 = -\tfrac{1}{2}(\alpha^2 + \gamma^2) g_0\qquad R_1 = \tfrac{1}{2} (\alpha^2 + \gamma^2) g_1\qquad R_2 = R_3 = 0~.
\end{equation}
As $R_2 = R_3 = 0$ is forced, what we find is a solution for $AdS_3 \times S^3 \times T^4$.

\subsubsection*{\texorpdfstring{$\mathbf{AdS_3 \times S^3 \times S^3 \times T^1}$}{AdS3 x S3 x S3 x T1}}\hypertarget{geom:AdS3S3S3T1}{}

The field equations admit the following solution:
\begin{equation}
  \begin{aligned}[m]
    \fGa &= \fF = 0\\
    \fG &= \beta_1 \nu_0 + \beta_2 \nu_1 + \beta_3 \nu_2\\
    \fH &= \alpha_1 \nu_0 + \alpha_2 \nu_1 + \alpha_3 \nu_2
  \end{aligned}
\end{equation}
with
\begin{equation}
  \alpha_1^2 = \alpha_2^2 + \alpha_3^2\qquad
	\beta_1^2 = \beta_2^2 + \beta_3^2\qquad
  \alpha_1\beta_2 = \beta_1\alpha_2\qquad
	\alpha_1\beta_3 = \beta_1\alpha_3\qquad
  \alpha_2\beta_3 = \beta_2\alpha_3~.
\end{equation}
The last three equations say that the vectors $(\alpha_1,\alpha_2,\alpha_3)$ and $(\beta_1,\beta_2,\beta_3)$ are collinear, so that $\fG$ and $\fH$ point in the same direction.  In this case we can then use the residual $\SO(2)$ duality transformations to set the $\beta_i=0$, whence we arrive at the simplified solution
\begin{equation}
  \begin{aligned}[m]
    \fGa &= \fG = \fF = 0\\
    \fH &= \alpha_1 \nu_0 + \alpha_2 \nu_1 + \alpha_3 \nu_2
  \end{aligned}
\end{equation}
with $\alpha_1^2 = \alpha_2^2 + \alpha_3^2$.  The Einstein equation then yields:
\begin{equation}
  R_0 = -\tfrac{1}{2}\alpha_1^2 g_0\qquad R_1 = \tfrac{1}{2}\alpha_2^2 g_1\qquad R_2 = \tfrac{1}{2}\alpha_3^2 g_2~,
\end{equation}
giving a solution for $AdS_3 \times S^3 \times S^3 \times T^1$.  This solution degenerates to one for $AdS_3 \times S^3 \times T^4$ whenever $\alpha_2 = 0$ or $\alpha_3 = 0$.

As mentioned briefly in Section \ref{sec:relat-with-symm}, to this background there corresponds a symmetric M-theory background with geometry $AdS_3 \times S^3 \times S^3 \times T^2$ and $F = \fH \wedge d\vartheta^2$.  This background is discussed in \cite[§4.6.3]{FigueroaO'Farrill:2011fj} and given in equation (64) in that paper, albeit in a somewhat different notation.

\subsubsection*{\texorpdfstring{$\mathbf{AdS_3 \times S^3 \times S^2 \times T^2}$}{AdS3 x S3 x S2 x T2}}\hypertarget{geom:AdS3S3S2T2}{}

The field equations admit the following solutions with $\xi_1, \xi_2 = \pm 1$:
\begin{enumerate}
\item
\begin{equation}
  \begin{aligned}[m]
    \fGa &= 0\\
    \fG &= \xi_1 \sqrt{\gamma_2^2 - \gamma_1^2} \nu_2 \wedge d\vartheta^2\\
    \fH &= \xi_2 \sqrt{\gamma_2^2 - \gamma_1^2} \nu_2 \wedge d\vartheta^1\\
    \fF &= (\gamma_1 \nu_0 + \gamma_2 \nu_1) \wedge \nu_2 - (\gamma_2 \nu_0 + \gamma_1 \nu_1) \wedge d\vartheta^1 \wedge d\vartheta^2~,
  \end{aligned}
\end{equation}
The Einstein equation then yields:
\begin{equation}
  R_0 = -\tfrac{1}{2}\gamma_2^2 g_0\qquad R_1 = \tfrac{1}{2}\gamma_1^2 g_1\qquad\text{and}\qquad R_2 = (\gamma_2^2 - \gamma_1^2) g_2~,
\end{equation}
giving a solution for the compact case $AdS_3 \times S^3 \times S^2 \times T^2$.  This solution degenerates to one for $AdS_3 \times S^3 \times T^4$ if $\gamma_1^2 = \gamma_2^2$, and to one for $AdS_3 \times S^2 \times T^5$ if $\gamma_1^2 = 0$.

\item Using the residual $\SO(2)$ duality, we can write a second solution as
\begin{equation}
  \begin{aligned}[m]
    \fGa &= \fG = 0\\
    \fH &= \alpha (\nu_0 + \xi_1 \nu_1)\\
    \fF &= \gamma (\nu_0 + \xi_1 \nu_1 ) \wedge (\nu_2 - \xi_1  d\vartheta^{12})~.
  \end{aligned}
\end{equation}
The Einstein equation then yields:
\begin{equation}
  R_0 = -\tfrac{1}{2}(\alpha^2 + \gamma^2) g_0\qquad R_1 = \tfrac{1}{2}(\alpha^2 + \gamma^2) g_1\qquad\text{and}\qquad R_2 = 0~.
\end{equation}
As $R_2 = 0$ is forced, what we obtain is a solution for $AdS_3 \times S^3 \times T^4$.
\end{enumerate}

\subsubsection*{\texorpdfstring{$\mathbf{AdS_3 \times S^3 \times T^4}$}{AdS3 x S3 x T4}}\hypertarget{geom:AdS3S3T4}{}

The field equations admit the following solution, where we have used the residual $\SO(2)$ duality transformation to set $\fG=0$ and where $d\vartheta^{ab} = d\vartheta^a \wedge d\vartheta^b$ and $\xi$ is a sign:
\begin{equation}  
  \begin{aligned}[m]  
   \fGa &= \fG = 0\\  
   \fH &= \alpha (\nu_0 + \xi \nu_1)\\  
   \fF &= \gamma (\nu_0 + \xi \nu_1) \wedge (d\vartheta^{12} - \xi d\vartheta^{34})~.
 \end{aligned}  
\end{equation}  
The Einstein equation then yields:
\begin{equation}  
  R_0 = -\tfrac{1}{2}(\alpha^2 + \gamma^2) g_0\qquad\text{and}\qquad R_1 = \frac{1}{2}(\alpha^2 + \gamma^2) g_1~,  
\end{equation}  
which is a solution for $AdS_3 \times S^3 \times T^4$.

The special case when $\gamma=0$ corresponds to a strongly symmetric background with only $\fH$ turned on.  The corresponding symmetric M-theory background has geometry $AdS_3 \times S^3 \times T^5$ and with $F=\alpha (\nu_0 + \xi \nu_1) \wedge d\vartheta^5$ and is discussed in \cite[§4.6.3]{FigueroaO'Farrill:2011fj} and particularly in equation (61) in that paper.

\subsubsection*{\texorpdfstring{$\mathbf{AdS_3 \times S^2 \times S^2 \times T^3}$}{AdS3 x S2 x S2 x T3}}\hypertarget{geom:AdS3S2S2T3}{}

This Ansatz has $(m_1,m_3,m_5^+) = (3,8,5)$, in the notation of Section \ref{sec:polynomial-systems}, whence a total of 24 parameters.  The resultant system of polynomials does not lend itself to symbolic solution; although we can exhibit an exact solution of the following form where $\xi$ is a sign:
\begin{equation}  
  \begin{aligned}[m]  
   \fGa &= 0\\  
   \fH &= \nu_1 \wedge (\alpha_1 d\vartheta^{2} + \alpha_2 d\vartheta^{3}) + \nu_2 \wedge (\alpha_3 d\vartheta^{2} + \alpha_4 d\vartheta^{3})\\ 
	 \fG &= \xi\left(\nu_1 \wedge (\alpha_2 d\vartheta^{2} - \alpha_1 d\vartheta^{3}) - \nu_2 \wedge (\alpha_4 d\vartheta^{2} - \alpha_3 d\vartheta^{3})\right)\\ 
   \fF &= \sqrt{\alpha_1^2  +\alpha_2^2 + \alpha_3^2 + \alpha_4^2} (\nu_0 \wedge d\vartheta^{23} - \nu_1 \wedge \nu_2 \wedge d\vartheta^{1})~,
 \end{aligned}  
\end{equation}
where $d\vartheta^{ab} = d\vartheta^a \wedge d\vartheta^b$.  The Einstein equation then yields:  
\begin{equation}  
  R_0 = -\tfrac{1}{2}(\alpha_1^2  +\alpha_2^2 + \alpha_3^2 + \alpha_4^2) g_0\qquad R_1 = (\alpha_1^2  +\alpha_2^2) g_1\qquad R_2 = (\alpha_3^2  +\alpha_4^2) g_2~,  
\end{equation} 
giving a solution for $AdS_2 \times S^2 \times S^2 \times T^3$.

\subsubsection*{\texorpdfstring{$\mathbf{AdS_3 \times S^2 \times T^5}$}{AdS3 x S2 x T5}}\hypertarget{geom:AdS3S2T5}{}

This Ansatz has $(m_1,m_3,m_5^+) = (5,16,11)$, for a total of 48 parameters.  The resultant system of polynomials does not lend itself to symbolic solution.  However we have seen that such backgrounds exist as limits of $AdS_3 \times S^3 \times S^2 \times T^2$.

\subsection{\texorpdfstring{$\mathbf{AdS_2}$}{AdS d=2} backgrounds}

The complexity of most $AdS_2$ backgrounds is such that for many of them we only have only partial results; that is, we find some exact solutions but we have been unable to determine the moduli space fully.

\subsubsection*{\texorpdfstring{$\mathbf{AdS_2 \times S^5 \times S^3}$ \bf{and} $\mathbf{AdS_2 \times SLAG_3 \times S^3}$}{AdS2 x S5 x S3 and AdS2 x SLAG3 x S3}}\hypertarget{geom:AdS2S5H3}{}

The field equations admit the following solution:
\begin{equation}
  \begin{aligned}[m]
    \fGa &= \fG = \fH = 0\\
    \fF &= \gamma (\nu_1 + \nu_0 \wedge \nu_2)~.
  \end{aligned}
\end{equation}
The Einstein equation yields:
\begin{equation}
  R_0 = -\tfrac14 \gamma^2 g_0\qquad R_1 = \tfrac14 \gamma^2 g_1\qquad R_2 = -\tfrac14 \gamma^2 g_2~,
\end{equation}
which yields a solution for $AdS_2 \times S^5 \times H^3$ and $AdS_2 \times SLAG_3 \times H^3$.  The existence of this background can be inferred from that of the $AdS_5 \times S^5$ and $AdS_4 \times SLAG_3$ backgrounds by virtue of the second observation in Section \ref{sec:further-observations}.

\subsubsection*{\texorpdfstring{$\mathbf{AdS_2 \times S^4 \times S^3 \times T^1}$}{AdS2 x S4 x S3 x T1}}\hypertarget{geom:AdS2S4S3T1}{}

The field equations first admit the following solution:
\begin{equation}
  \begin{aligned}[m]
    \fGa &= \kappa d\vartheta\\
    \fG &= \fF = 0\\
    \fH &= \frac{\xi_1}{\sqrt2} \kappa \left(\nu_0 \wedge d\vartheta + \xi_2 \sqrt5 \nu_2 \right)~,
  \end{aligned}
\end{equation}
where $\vartheta$ is a flat coordinate on $T^1$ and $\xi_{1,2}=\pm 1$.  The Einstein equation yields:
\begin{equation}
  R_0 = -\tfrac12\kappa^2 g_0\qquad R_1 = -\tfrac14\kappa^2 g_1\qquad R_2 = \kappa^2 g_2~,
\end{equation}
giving a solution for $AdS_2 \times H^4 \times S^3 \times T^1$.

The field equations also admit the following solution, again with $\xi_{1,2}=\pm 1$:
\begin{equation}
  \begin{aligned}[m]
    \fGa &= \alpha d\vartheta\\
    \fG &= \xi_1 \sqrt2 \alpha \nu_0 \wedge d\vartheta\\
    \fH &= 0\\
    \fF &= \xi_2 \alpha (\nu_0 \wedge \nu_2 - \nu_1 \wedge d\vartheta)~.
  \end{aligned}
\end{equation}
The Einstein equation yields:
\begin{equation}
  R_0 = - \alpha^2 g_0\qquad R_1 = \half \alpha^2 g_1\qquad R_2 = 0~.
\end{equation}
As $R_2 = 0$ is forced, what we obtain is a solution for $AdS_2 \times S^4 \times T^4$.

\subsubsection*{\texorpdfstring{$\mathbf{AdS_2 \times S^5 \times S^2 \times T^1}$ \bf{and} $\mathbf{AdS_2 \times SLAG_3 \times S^2 \times T^1}$}{AdS2 x S5 x S2 x T1 and AdS2 x SLAG3 x S2 x T1}}\hypertarget{geom:AdS2S5S2T1}{}

The field equations admit the following solution with $\xi_1,\xi_2 = \pm 1$:
\begin{equation}
  \begin{aligned}[m]
    \fGa &= \sqrt{3}\kappa d\vartheta\\
    \fG &= \sqrt{5} \left(\xi_1\sqrt{\alpha^2+\kappa^2} \nu_0 \wedge d\vartheta + \xi_2 \alpha \nu_2 \wedge d\vartheta\right)\\
    \fH &= \alpha \nu_0 \wedge d\vartheta + \xi_1 \xi_2 \sqrt{\alpha^2+\kappa^2} \nu_2 \wedge d\vartheta\\
    \fF &= 0~.
  \end{aligned}
\end{equation}
The Einstein equation yields:
\begin{equation}
  R_0 = -(3\alpha^2 + 2\kappa^2) g_0\qquad R_1 = \tfrac{1}{2}\kappa^2 g_1\qquad R_2 = (3 \alpha^2 + \kappa^2) g_2\qquad R_3 = 0~,
\end{equation}
giving a solution for $AdS_2 \times S^5 \times S^2 \times T^1$ and $AdS_2 \times SLAG_3 \times S^2 \times T^1$.  This solution degenerates to one for $AdS_2 \times S^2 \times T^6$ when $\kappa=0$.

\subsubsection*{\texorpdfstring{$\mathbf{AdS_2 \times S^3 \times S^3 \times S^2}$}{AdS2 x S3 x S3 x S2}}\hypertarget{geom:AdS2S3H3S2}{}

The field equations admit the following solution:
\begin{equation}
  \begin{aligned}[m]
    \fGa &= \fG = \fH = 0\\
    \fF &= \nu_0 \wedge(\gamma_1 \nu_1 + \gamma_2 \nu_2 ) + (\gamma_2 \nu_1 - \gamma_1 \nu_2) \wedge \nu_3~.
  \end{aligned}
\end{equation}
The Einstein equation yields:
\begin{equation}
  R_0 = -\tfrac{1}{4}(\gamma_1^2 + \gamma_2^2) g_0\qquad R_1 = -\tfrac{1}{4}(\gamma_1^2 - \gamma_2^2) g_1\qquad R_2 = \tfrac{1}{4}(\gamma_1^2 - \gamma_2^2) g_2\qquad R_3 = \tfrac{1}{4}(\gamma_1^2 + \gamma_2^2) g_3~,
\end{equation}
giving a solution for $AdS_2 \times S^3 \times H^3 \times S^2$.  This solution degenerates to $AdS_2 \times S^2 \times T^6$ whenever $\gamma_1^2 = \gamma_2^2$.

\subsubsection*{\texorpdfstring{$\mathbf{AdS_2 \times \mathbb{CP}^2 \times S^3 \times T^1}$}{AdS2 x CP2 x S3 x T1}}\hypertarget{geom:AdS2CP2S3T1}{}

The field equations admit the following five branches of solutions, with $\xi_i = \pm 1$:
\begin{enumerate}
\item
\begin{equation}
  \begin{aligned}[m]
    \fGa &= \fG = \fF = 0\\
    \fH &= \alpha (\xi_1 \sqrt{2} \nu_0 + \omega_{\mathbb{CP}^2}) \wedge d\vartheta~.
  \end{aligned}
\end{equation}
The Einstein equation yields:
\begin{equation}
  R_0 = -\alpha^2 g_0\qquad R_1 = \tfrac{1}{2}\alpha^2 g_1\qquad R_2 = 0~.
\end{equation}
Since $R_2=0$, what we have found is a background with geometry $AdS_2 \times \mathbb{CP}^2 \times T^4$.

\item
\begin{equation}
  \begin{aligned}[m]
    \fGa &= \xi_1 \sqrt{6(\alpha^2-\beta^2)} d\vartheta\\
		\fG &= \xi_2 \sqrt{10}\alpha \nu_0 \wedge d\vartheta + \frac{\beta}{\sqrt{5}} \omega_{\mathbb{CP}^2} \wedge d\vartheta\\
		\fH &= \xi_2 \sqrt{2}\beta \nu_0 \wedge d\vartheta + \alpha \omega_{\mathbb{CP}^2} \wedge d\vartheta\\
    \fF &= 0~.
  \end{aligned}
\end{equation}
The Einstein equation yields:
\begin{equation}
  R_0 = -2(2\alpha^2 + \beta^2) g_0\qquad R_1 = \tfrac{3}{2}(\alpha^2 + \beta^2) g_1\qquad R_2 = (\alpha^2 - \beta^2)g_2~,
\end{equation}
giving a solution for $AdS_2 \times \mathbb{CP}^2 \times S^3 \times T^1$.

\item
\begin{equation}
  \begin{aligned}[m]
    \fGa &= \fG = \fH = 0\\
    \fF &= \gamma (\omega_{\mathbb{CP}^2}+ \xi \sqrt2 \nu_0) \wedge (\omega_{\mathbb{CP}^2} - \xi \sqrt2 \nu_2)~.
  \end{aligned}
\end{equation}
The Einstein equation yields:
\begin{equation}
  R_0 = -2 \gamma^2 g_0\qquad R_1 = \gamma^2 g_1\qquad R_2 = 0~.
\end{equation}
Since $R_2=0$, this is actually a solution for $AdS_2 \times \mathbb{CP}^2 \times T^4$.

\item
\begin{equation}
  \begin{aligned}[m]
    \fGa &= \xi_1 \sqrt{2} \alpha_1 d\vartheta\\
		\fG &= \fF = 0\\
		\fH &= \frac{1}{\sqrt{5}}\alpha_1 \nu_2 + (\xi_2 \sqrt{\alpha_1^2 + 2\alpha_2^2} \nu_0 + \alpha_2 \omega_{\mathbb{CP}^2}) \wedge d\vartheta~.
  \end{aligned}
\end{equation}
The Einstein equation yields:
\begin{equation}
  R_0 = -(\alpha_1^2 + \alpha_2^2) g_0\qquad R_1 = \tfrac{1}{2}(\alpha_2^2 - \alpha_1^2) g_1\qquad R_2 = 2 \alpha_1^2 g_2~,
\end{equation}
giving a solution for $AdS_2 \times \mathbb{CP}^2 \times S^3 \times T^1$ and $AdS_2 \times \mathbb{CH}^2 \times S^3 \times T^1$, depending on the sign of $\alpha_2^2 - \alpha_1^2$.  When $\alpha_1^2 = \alpha_2^2$ we get a solution for $AdS_2 \times S^3 \times T^5$.

\item
\begin{equation}
  \begin{aligned}[m]
    \fGa &= \xi_1 \beta d\vartheta\\
		\fG &= \frac{1}{\sqrt{2}}\beta \nu_0 \wedge d\vartheta\\
		\fH &= 0\\
    \fF &= \xi_2 \beta(\nu_1 \wedge d\vartheta - \nu_0 \wedge \nu_2)~.
  \end{aligned}
\end{equation}
The Einstein equation yields:
\begin{equation}
  R_0 = -\beta^2 g_0\qquad R_1 = \tfrac{1}{2}\beta^2 g_1\qquad R_2 = 0~.
\end{equation}
Since $R_2=0$, what we find is a background with geometry $AdS_2 \times \mathbb{CP}^2 \times T^4$.
\end{enumerate}

\subsubsection*{\texorpdfstring{$\mathbf{AdS_2 \times S^3 \times S^3 \times T^2}$}{AdS2 x S3 x S3 x T2}}\hypertarget{geom:AdS2S3S3T2}{}

The field equations admit the following solution with $\xi_1,\xi_2 = \pm 1$:
\begin{equation}
  \begin{aligned}[m]
    \fGa &= 0\\
    \fG &= \xi_1 \sqrt{\gamma_1^2 + \gamma_2^2} \nu_0 \wedge d\vartheta^1\\
    \fH &= \xi_2 \sqrt{\gamma_1^2 + \gamma_2^2} \nu_0 \wedge d\vartheta^2\\
    \fF &= \nu_0 \wedge(\gamma_2 \nu_1 + \gamma_1 \nu_2) + (\gamma_1 \nu_1 - \gamma_2 \nu_2) \wedge d\vartheta^{12}~.
  \end{aligned}
\end{equation}
The Einstein equation then yields:
\begin{equation}
  R_0 = -(\gamma_1^2 + \gamma_2^2) g_0 \qquad R_1 = \tfrac{1}{2} \gamma_1^2 g_1 \qquad R_2 = \tfrac{1}{2} \gamma_2^2 g_2~,
\end{equation}
which gives a solution for $AdS_2 \times S^3 \times S^3 \times T^2$.  This solution degenerates to one for $AdS_2 \times S^3 \times T^5$ whenever $\gamma_1 = 0$ or $\gamma_2 = 0$.

\subsubsection*{\texorpdfstring{$\mathbf{AdS_2 \times G_{\mathbb{R}}^{+}(2,5) \times T^2}$ \bf{and} $\mathbf{AdS_2 \times \mathbb{CP}^3 \times T^2}$}{AdS2 x G[R+](2,5) x T2 and AdS2 x CP3 x T2}}\hypertarget{geom:AdS2CP3T2}{}

In the strongly symmetric case, the field equations admit the following solution with $\xi = \pm 1$:
\begin{equation}
  \begin{aligned}[m]
    \fGa &= \fG = \fF = 0\\
    \fH &= \alpha(\xi \sqrt3 \nu_0 + \omega) \wedge d\vartheta^1~,
  \end{aligned}
\end{equation}
with $\omega$ the Kähler form in the relevant hermitian symmetric space $\CP^3$ or $G_\RR^+(2,5)$.  The Einstein equation then yields:
\begin{equation}
  R_0 = -\tfrac{1}{2}\alpha^2 g_0\qquad\text{and}\qquad R_1 = \tfrac{1}{6}\alpha^2 g_1~,
\end{equation}
giving a solution for $AdS_2 \times G_\RR^+(2,5) \times T^2$ and $AdS_2 \times \mathbb{CP}^3 \times T^2$.

These backgrounds, once reinterpreted as symmetric IIA backgrounds, lift to  symmetric M-theory backgrounds with geometries $AdS_2 \times G_\RR^+(2,5) \times T^3$ and $AdS_2 \times \CP^3 \times T^3$ and $F=\fH \wedge d\vartheta^3$ in both cases.  These backgrounds are discussed in \cite[§4.7.6]{FigueroaO'Farrill:2011fj} and given in equation (106) in that paper.

When $\fGa\neq 0$, the field equations admit the following set of related solutions with $\xi_i=\pm 1$:
\begin{enumerate}
\item
\begin{equation}
  \begin{aligned}[m]
    \fGa &= \kappa d\vartheta^1\\
    \fH &= \frac{\kappa}{\sqrt2} (\xi_2 \sqrt3 \nu_0 \wedge d\vartheta^1 + \xi_3 \omega \wedge d\vartheta^2)\\
		\fG &= \xi_1 \sqrt2 \kappa \nu_0 \wedge d\vartheta^2\\
		\fF &= 0~,
  \end{aligned}
\end{equation}
The Einstein equation then yields:
\begin{equation}
  R_0 = -\tfrac{3}{2}\kappa^2 g_0\qquad\text{and}\qquad R_1 = \tfrac{1}{2}\kappa^2 g_1~,
\end{equation}

\item
\begin{equation}
  \begin{aligned}[m]
    \fGa &= \kappa d\vartheta^1\\
    \fH &= \frac{\kappa}{\sqrt2} (\xi_2 \nu_0 \wedge d\vartheta^2 + \xi_3 \frac{1}{\sqrt3}\omega \wedge d\vartheta^1)\\
		\fG &= \xi_1 \sqrt2 \kappa \nu_0 \wedge d\vartheta^1\\
		\fF &= 0~,
  \end{aligned}
\end{equation}
The Einstein equation then yields:
\begin{equation}
  R_0 = -\kappa^2 g_0\qquad\text{and}\qquad R_1 = \tfrac{1}{3}\kappa^2 g_1~,
\end{equation}

\item
\begin{equation}
  \begin{aligned}[m]
    \fGa &= \kappa d\vartheta^1\\
    \fH &= 0\\
		\fG &= \xi_1 \sqrt2 \kappa \nu_0 \wedge d\vartheta^2\\
		\fF &= \xi_2\kappa(\nu_0 \wedge \omega \wedge d\vartheta^1 - \nu_1 \wedge d\vartheta^2)~,
  \end{aligned}
\end{equation}
The Einstein equation then yields:
\begin{equation}
  R_0 = -\tfrac{3}{2}\kappa^2 g_0\qquad\text{and}\qquad R_1 = \tfrac{1}{2}\kappa^2 g_1~,
\end{equation}

\item
\begin{equation}
  \begin{aligned}[m]
    \fGa &= \kappa d\vartheta^1\\
    \fH &= 0\\
		\fG &= \xi_1 \sqrt2 \kappa \nu_0 \wedge d\vartheta^1\\
		\fF &= \xi_2\frac{\kappa}{\sqrt3}(\nu_0 \wedge \omega \wedge d\vartheta^2 + \nu_1 \wedge d\vartheta^1)~,
  \end{aligned}
\end{equation}
The Einstein equation then yields:
\begin{equation}
  R_0 = -\kappa^2 g_0\qquad\text{and}\qquad R_1 = \tfrac{1}{3}\kappa^2 g_1~,
\end{equation}

\end{enumerate}
giving solutions for $AdS_2 \times G_{\mathbb{R}}^{+}(2,5) \times T^2$ and $AdS_2 \times \mathbb{CP}^3 \times T^2$.

\subsubsection*{\texorpdfstring{$\mathbf{AdS_2 \times S^4 \times S^2 \times T^2}$}{AdS2 x S4 x S2 x T2}}\hypertarget{geom:AdS2S4S2T2}{}

This Ansatz has $(m_1,m_3,m_5^+) = (2,4,2)$, for a total of 12 parameters.  Let $\nu_0,\nu_1,\nu_2$ denote the volume forms of $AdS_2$, $S^4$ and $S^2$, respectively.   There are several branches, which we now enumerate.  We have omitted certain branches where the one of the spheres is forced to be flat.

\begin{enumerate}
\item With $\xi_{1,2}=\pm 1$, we have
  \begin{equation}
    \begin{aligned}[m]
      \fGa &= \alpha d\vartheta^2\\
      \fG &= \left(\beta \nu_2 + \xi_1 \sqrt{2\alpha^2 + \beta^2} \nu_0 \right) \wedge d\vartheta^2 \\
      \fH &= 0\\
      \fF &= \xi_2 \alpha \left(\nu_0\wedge\nu_2 \wedge d\vartheta^1 - \nu_1 \wedge d\vartheta^2\right)~,    
    \end{aligned}
  \end{equation}
  which gives rise to the following Ricci curvatures:
  \begin{equation}
    R_0 = - \half (2\alpha^2 + \beta^2) g_0 \qquad R_1 = \half \alpha^2 g_1 \qquad R_2 = \half \beta^2 g_2~.
  \end{equation}
  Therefore we obtain a solution for $AdS_2 \times S^4 \times S^2 \times T^2$ in the generic case, or $AdS_2 \times S^2 \times T^6$ if $\alpha=0$ and $AdS_2 \times S^4 \times T^4$ if $\beta = 0$.  By the second observation in Section~\ref{sec:further-observations}, we obtain solutions for $AdS_2 \times S^2 \times S^2 \times S^2 \times T^2$ and $AdS_2 \times S^2 \times S^2 \times T^4$.

\item Again with $\xi_{1,2}$ a sign, we have
  \begin{equation}
    \begin{aligned}[m]
      \fGa &= \alpha d\vartheta^1\\
      \fG &= \left(\beta \nu_2 + \xi_1 \sqrt{2\alpha^2 + \beta^2} \nu_0 \right) \wedge d\vartheta^2 \\
      \fH &= 0\\
      \fF &= \xi_2 \sqrt{3} \alpha \left(\nu_0\wedge\nu_2 \wedge d\vartheta^1 - \nu_1 \wedge d\vartheta^2\right)~,
    \end{aligned}
  \end{equation}
  which gives rise to the following Ricci curvatures:
  \begin{equation}
    R_0 = - \half (3 \alpha^2 + \beta^2) g_0 \qquad R_1 = \alpha^2 g_1 \qquad R_2 = \half (\beta^2 - \alpha^2) g_2~.
  \end{equation}
  Therefore we obtain a solution for $AdS_2 \times S^4 \times S^2 \times T^2$ for $|\beta|>|\alpha|$, $AdS_2 \times S^4 \times T^4$ for $|\beta| = |\alpha|$ and $AdS_2 \times S^4 \times H^2 \times T^2$ for $|\beta|<|\alpha|$.  Again this also gives solutions for $AdS_2 \times S^2 \times S^2 \times S^2 \times T^2$, $AdS_2 \times S^2 \times S^2 \times T^4$ and $AdS_2 \times S^2 \times S^2 \times H^2 \times T^2$.


\item With $\xi_{1,2,3}=\pm 1$, we have
  \begin{equation}
    \begin{aligned}[m]
      \fGa &= \alpha d\vartheta^1\\
      \fG &= \xi_3\sqrt{\beta^2+2\alpha^2}\nu_0 \wedge d\vartheta^2 + \beta \nu_2 \wedge d\vartheta^1\\
      \fH &= \sqrt{\beta^2+\tfrac{3}{2}\alpha^2}(\xi_1 \nu_0 \wedge d\vartheta^1 + \xi_2 \nu_2 \wedge d\vartheta^2)\\
      \fF &= 0~,
    \end{aligned}
  \end{equation}
  which gives rise to the following Ricci curvatures:
  \begin{equation}
    R_0 = - (\tfrac{3}{2}\alpha^2 + \beta^2) g_0 \qquad R_1 = \tfrac{1}{4}\alpha^2 g_1 \qquad R_2 = (\alpha^2 + \beta^2) g_2~.
  \end{equation}
  Therefore we obtain a solution for $AdS_2 \times S^4 \times S^2 \times T^2$.

\item Again with $\xi_{1,2,3}$ a sign, we have
  \begin{equation}
    \begin{aligned}[m]
      \fGa &= \alpha d\vartheta^1\\
      \fG &= \xi_3\sqrt{\beta^2+2\alpha^2}\nu_0 \wedge d\vartheta^1 + \beta \nu_2 \wedge d\vartheta^2\\
      \fH &= \sqrt{\beta^2+\tfrac{1}{2}\alpha^2}(\xi_1 \nu_0 \wedge d\vartheta^2 + \xi_2 \nu_2 \wedge d\vartheta^1)\\
      \fF &= 0~,
    \end{aligned}
  \end{equation}
  which gives rise to the following Ricci curvatures:
  \begin{equation}
    R_0 = - (\alpha^2 + \beta^2) g_0 \qquad R_1 = \tfrac{1}{4}\alpha^2 g_1 \qquad R_2 = (\tfrac{1}{2}\alpha^2 + \beta^2) g_2~.
  \end{equation}
  Therefore we obtain a solution for $AdS_2 \times S^4 \times S^2 \times T^2$.

\item Again with $\xi_{1,2,3}$ a sign, we have
  \begin{equation}
    \begin{aligned}[m]
      \fGa &= \alpha d\vartheta^1\\
      \fG &= \beta(\nu_2 \wedge d\vartheta^1 - \xi_2 \xi_3 \nu_0 \wedge d\vartheta^2) + \sqrt{2} \xi_1 \xi_3 \alpha \nu_0 \wedge d\vartheta^1\\
      \fH &= \xi_1 \beta(\nu_0 \wedge d\vartheta^1 - \xi_2 \xi_3 \nu_2 \wedge d\vartheta^2) + \tfrac{1}{\sqrt{2}}\alpha(\xi_3 \nu_2 \wedge d\vartheta^1 + \xi_2 \nu_0 \wedge d\vartheta^2)\\
      \fF &= 0~,
    \end{aligned}
  \end{equation}
  which gives rise to the following Ricci curvatures:
  \begin{equation}
    R_0 = - (\alpha^2 + \beta^2) g_0 \qquad R_1 = \tfrac{1}{4}\alpha^2 g_1 \qquad R_2 = (\tfrac{1}{2}\alpha^2 + \beta^2) g_2~.
  \end{equation}
  Therefore we obtain a solution for $AdS_2 \times S^4 \times S^2 \times T^2$.

\item Again with $\xi_{1,2,3}$ a sign, we have
  \begin{equation}
    \begin{aligned}[m]
      \fGa &= \alpha d\vartheta^1\\
      \fG &= \beta(\nu_2 \wedge d\vartheta^2 - 3\xi_2 \xi_3 \nu_0 \wedge d\vartheta^1) + \xi_1 \xi_3 \sqrt{2(\alpha^2-4\beta^2)}\nu_0 \wedge d\vartheta^2\\
      \fH &= \xi_1 \sqrt{\tfrac{3}{2}(\alpha^2 - 4\beta^2)}(\nu_0 \wedge d\vartheta^1 -\xi_2\xi_3\nu_2 \wedge d\vartheta^2) + \sqrt{3}\beta(\xi_2 \nu_0 \wedge d\vartheta^2 + \xi_3 \nu_2 \wedge d\vartheta^1)\\
      \fF &= 0~,
    \end{aligned}
  \end{equation}
  which gives rise to the following Ricci curvatures:
  \begin{equation}
    R_0 = -(\tfrac{3}{2}\alpha^2 - \beta^2) g_0 \qquad R_1 = \tfrac{1}{4}\alpha^2 g_1 \qquad R_2 = (\alpha^2 - \beta^2) g_2~.
  \end{equation}
  Therefore we obtain a solution for $AdS_2 \times S^4 \times S^2 \times T^2$.
\end{enumerate}

There is an additional branch which does not seem to be explicitly parametrisable, in the sense that the equations are not solvable in terms of radicals.

\subsubsection*{\texorpdfstring{$\mathbf{AdS_2 \times S^3 \times S^2 \times S^2 \times T^1}$}{AdS2 x S3 x S2 x S2 x T1}}\hypertarget{geom:AdS2S3S2S2T1}{}

This Ansatz has $(m_1,m_3,m_5^+) = (1,4,3)$, for a total of 12 parameters. The resultant system of polynomials does not lend itself to symbolic solution but an exact background with geometry $AdS_2 \times S^3 \times S^2 \times H^2 \times T^1$ can be written down from the background found for \hyperlink{geom:AdS4S3S2T1}{$AdS_4 \times S^3 \times S^2 \times T^1$}.
In addition, we can exhibit an exact solution of the following form with $\xi_i=\pm 1$:
\begin{equation}  
  \begin{aligned}[m]  
   \fGa &= \sqrt{2}\kappa d\vartheta\\  
   \fH &= \xi_1 \sqrt{5} \kappa \nu_1 + \left(\xi_2 \sqrt{\alpha^2 + \beta^2 + \kappa^2} \nu_0 + \alpha \nu_2 + \beta \nu_3\right) \wedge d\vartheta\\ 
	 \fG &= \fF = 0~,
 \end{aligned}  
\end{equation}
The Einstein equation then yields:  
\begin{equation}  
  R_0 = -\tfrac{1}{2}(2\kappa^2 + \alpha^2 + \beta^2) g_0\qquad R_1 = 2\kappa^2 g_1\qquad R_2 = \tfrac{1}{2}(\alpha^2 - \kappa^2) g_2\qquad R_3 = \tfrac{1}{2}(\beta^2 - \kappa^2) g_3~,  
\end{equation} 
giving solutions for $AdS_2 \times S^3 \times S^2 \times S^2 \times T^1$, $AdS_2 \times S^3 \times H^2 \times S^2 \times T^1$, and $AdS_2 \times S^3 \times H^2 \times H^2 \times T^1$.
This solution degenerates to one for $AdS_2 \times S^2 \times S^2 \times T^4$ whenever $\kappa = 0$, $AdS_2 \times S^2 \times T^6$ whenever $\kappa = \alpha = 0$ or $\kappa = \beta = 0$, and $AdS_2 \times S^3 \times H^2 \times T^2$ whenever $\alpha^2 = \kappa^2$ with $\beta^2 < \kappa^2$ or $\beta^2 = \kappa^2$ with $\alpha^2 < \kappa^2$.

\subsubsection*{\texorpdfstring{$\mathbf{AdS_2 \times \mathbb{CP}^2 \times S^2 \times T^2}$}{AdS2 x CP2 x S2 x T2}}\hypertarget{geom:AdS2CP2S2T2}{}

This Ansatz has $(m_1,m_3,m_5^+) = (2,6,5)$, for a total of 19 parameters.  The resultant system of polynomials does not lend itself to symbolic solution.  Numerical optimization suggests that solutions exist for both $AdS_2 \times \mathbb{CP}^2 \times S^2 \times T^2$ and $AdS_2 \times \mathbb{CP}^2 \times H^2 \times T^2$.  We can exhibit an exact solution of the following form:
\begin{equation}  
  \begin{aligned}[m]  
   \fGa &= \kappa(d\vartheta^{1} + d\vartheta^{2})\\
   \fH &= \tfrac{\sqrt{7}}{2}\kappa(\nu_0 \wedge d\vartheta^{1} + \nu_2 \wedge d\vartheta^{2})\\ 
	 \fG &= \tfrac{1}{2}\kappa(\nu_0 \wedge (d\vartheta^{1} + 4 d\vartheta^{2}) + \nu_2 \wedge d\vartheta^{2})\\ 
   \fF &= 0~,
 \end{aligned}  
\end{equation}
The Einstein equation then yields:  
\begin{equation}  
  R_0 = -\tfrac{5}{2}\kappa^2 g_0\qquad R_1 = \tfrac{1}{2}\kappa^2 g_1\qquad R_2 = \tfrac{3}{2}\kappa^2 g_2~,  
\end{equation} 
giving a solution for $AdS_2 \times \mathbb{CP}^2 \times S^2 \times T^2$.  Notice that the solution does not depend on any of the invariant forms of $\CP^2$, whence it also gives a solution for $AdS_2 \times X^4 \times S^2 \times T^2$, where $X$ is any compact (since the curvature is positive) four-dimensional riemannian symmetric space: $S^4$, $\CP^2$ or $S^2 \times S^2$.  In particular, this solution belongs to the branch of \hyperlink{geom:AdS2S4S2T2}{$AdS_2 \times S^4 \times S^2 \times T^2$} with $\fF=0$ and $\fG\neq 0$ whose general solution cannot be expressed in terms of radicals.

We can also exhibit an exact solution of the following form with $\xi_{1,2}=\pm 1$:
\begin{equation}  
  \begin{aligned}[m]  
   \fGa &= \fH = 0\\
	 \fG &=\xi_1\nu_0 \wedge (\gamma_1 d\vartheta^{1} - \gamma_2 d\vartheta^{2}) + \tfrac{1}{\sqrt{2}}\xi_2\nu_1 \wedge (\gamma_2 d\vartheta^{1} + \gamma_1 d\vartheta^{2})\\ 
   \fF &= \sqrt{2} (\nu_1 \wedge (\gamma_1 d\vartheta^{1} - \gamma_2 d\vartheta^{2}) + \nu_0 \wedge \nu_2 \wedge (\gamma_2 d\vartheta^{1} + \gamma_1 d\vartheta^{2}))~,
 \end{aligned}  
\end{equation}
The Einstein equation then yields:  
\begin{equation}
  R_0 = -(\gamma_1^2 + \gamma_2^2) g_0\qquad R_1 = \tfrac{3}{4}(\gamma_1^2 + \gamma_2^2) g_1\qquad R_2 = -\tfrac{1}{2}(\gamma_1^2 + \gamma_2^2) g_2~,  
\end{equation} 
giving a solution for $AdS_2 \times \mathbb{CP}^2 \times H^2 \times T^2$.

\subsubsection*{\texorpdfstring{$\mathbf{AdS_2 \times S^4 \times T^4}$}{AdS2 x S4 x T4}}\hypertarget{geom:AdS2S4T4}{}

We have already found such backgrounds when studying the geometries $AdS_2 \times S^4 \times S^3 \times T^1$ and $\AdS_2 \times S^4 \times S^2 \times T^2$, but in fact we can solve for the moduli space exactly and we find an additional branch.  The invariant forms are the volume form $\nu$ for $AdS_2$, the volume form $\sigma$ for $S^4$ and any constant-coefficient form on $T^4$.  We will let $d\vartheta^i$, $i=1,2,3,4$, denote an orthonormal coframe for $T^4$ and $\tau = d\vartheta^{1234}$ the corresponding volume form.

The most general Ansatz for a symmetric background is given by
\begin{equation}
  \begin{aligned}[m]
    \fGa &= \alpha \\
    \fG &= \nu \wedge \beta + \star\gamma\\
    \fH &= \nu \wedge \beta' + \star\gamma'\\
    \fF &= \nu \wedge \star\delta + \sigma\wedge \delta~,
  \end{aligned}
\end{equation}
where $\alpha,\beta,\beta',\gamma,\gamma',\delta$ are invariant 1-forms on $T^4$.  The field equations \eqref{eq:IIBEoMsym} become
\begin{equation}
  \label{eq:AdS2S4T4MEqns}
  \begin{aligned}[m]
    -|\beta'|^2 + |\gamma'|^2 &= -|\beta|^2 + |\gamma|^2 + 2|\alpha|^2\\
    0&= -\left<\beta,\beta'\right> + \left<\gamma,\gamma'\right>\\
    0&= \beta' \wedge \delta\\
    0&= \left<\delta,\gamma'\right>\\
    0&= \beta \wedge \delta\\
    0&= \left<\delta,\gamma\right>\\
    0&= \left<\beta,\gamma'\right> - \left<\beta',\gamma\right>
  \end{aligned}
\end{equation}
together with the Einstein equation, of which the $T^4$ components read
\begin{multline}
  \label{eq:AdS2S4T4EET4}
  0= \half\alpha_i\alpha_j - \half \beta_i\beta_j - \half \beta'_i\beta'_j - \half \gamma_i\gamma_j - \half \gamma'_i\gamma'_j + \half \delta_i\delta_j\\
  + \delta_{ij} \left(\tfrac18 |\beta|^2 + \tfrac18 |\beta'|^2 +\tfrac38 |\gamma|^2 + \tfrac38 |\gamma'|^2 - \tfrac14 |\delta|^2\right)~.
\end{multline}
We first show that $\delta\neq 0$.  Indeed, tracing the above equation we see that
\begin{equation}
  \half |\delta|^2 = \half |\alpha|^2 + |\gamma|^2 + |\gamma'|^2~,
\end{equation}
whence if $\delta=0$, so are $\alpha,\gamma,\gamma'$.  Two of the remaining equations for $\beta$ and $\beta'$ are then $|\beta|^2 = |\beta'|^2$ and $\left<\beta,\beta'\right>=0$.  Using the $SO(4)$ symmetry of $T^4$ we can choose $\beta = \beta_1 d\vartheta^1$ and $\beta' = \beta'_2 d\vartheta^2$ with $\beta_1^2 = (\beta'_2)^2$.  Then the $(33)$ component of equation \eqref{eq:AdS2S4T4EET4} says that $|\beta|^2 + |\beta'|^2 = 0$, whence $\beta = \beta'=0$, contradicting the fact that the geometry is not Ricci-flat.  Therefore $\delta \neq 0$.

Using the $\SO(4)$ symmetry we may set $\delta = \delta_1 d\vartheta^1$, with $\delta_1\neq 0$, and since $\beta \wedge \delta =0=\beta'\wedge \delta$, also $\beta = \beta_1 d\vartheta^1$ and $\beta'=\beta'_1 d\vartheta^1$.  Since $\gamma$ and $\gamma'$ are perpendicular to $\delta$, we can use the stabilising $SO(3)$ to set $\gamma = \gamma_2d\vartheta^2$ and then the stabilising $SO(2)$ to set $\gamma' = \gamma'_2 d\vartheta^2 + \gamma'_3 d\vartheta^3$, whereas $\alpha$ remains arbitrary.  The $(14)$, $(24)$ and $(34)$ components of equation \eqref{eq:AdS2S4T4EET4} give
\begin{equation}
  \alpha_1\alpha_4 = \alpha_2\alpha_4 = \alpha_3\alpha_4 = 0~,
\end{equation}
whence we have two branches to consider:
\begin{enumerate}
\item First branch: $\alpha_4\neq 0$, whence $\alpha_1=\alpha_2=\alpha_3=0$.  Equations \eqref{eq:AdS2S4T4MEqns} become simply
  \begin{equation}
    \label{eq:MEBr1}
    \gamma_2\gamma'_2 = \beta_1\beta'_1 \qquad\text{and}\qquad - (\beta'_1)^2 + (\gamma'_2)^2 + (\gamma'_3)^2 = -\beta_1^2 + \gamma_2^2 + 2 \alpha_4^2~,
  \end{equation}
  whereas equations \eqref{eq:AdS2S4T4EET4} now become
  \begin{equation}
    \begin{aligned}[m]
      0 &= \gamma'_2 \gamma'_3\\
      0 &= \beta_1^2 + (\beta'_1)^2 + 3 \gamma_2^2 + 3 (\gamma'_2)^2 - (\gamma'_3)^2 - 2\delta_1^2\\
      0 &= \beta_1^2 + (\beta'_1)^2 - \gamma_2^2 - (\gamma'_2)^2 + 3 (\gamma'_3)^2 -  2 \delta_1^2\\
      0 &= \beta_1^2 + (\beta'_1)^2 - \gamma_2^2 - (\gamma'_2)^2 - (\gamma'_3)^2 - \tfrac23 \delta_1^2\\
      0 &= 4 \alpha_4^2 + \beta_1^2 + (\beta'_1)^2 + 3\gamma_2^2 + 3(\gamma'_2)^2 + 3 (\gamma'_3)^2 -  2 \delta_1^2~.
    \end{aligned}
  \end{equation}
  Subtracting the second of the above equations from the last, we find that $\alpha_4 = \gamma'_3=0$.  Subtracting the third from the last we now find $\gamma_2=\gamma'_2=0$.  Finally subtracting the next to last equation from the last equation that $\delta_1=0$, which is a contradiction.
  
\item Second branch: $\alpha_4 = 0$.  Then the $(44)$ component of equation \eqref{eq:AdS2S4T4EET4} says that the term multiplying $\delta_{ij}$ vanishes separately, whence the resulting equations are now
  \begin{equation}
    \begin{aligned}[m]
      0&= \alpha_1\alpha_2\\
      0&= \alpha_1\alpha_3\\
      0&= \alpha_2\alpha_3 - \gamma'_2 \gamma'_3\\ 
      0&= \gamma_2\gamma'_2 - \beta_1\beta'_1\\
      0&= (\beta'_1)^2 - (\gamma'_2)^2 - (\gamma'_3)^2 - \beta_1^2 + \gamma_2^2 +2 \alpha_1^2 + 2 \alpha_2^2 + 2 \alpha_3^2\\
      0 &= 4\alpha_3^2 + \beta_1^2 + (\beta'_1)^2 + 3 \gamma_2^2 + 3 (\gamma'_2)^2 - (\gamma'_3)^2 - 2\delta_1^2\\
      0 &= 4\alpha_2^2 + \beta_1^2 + (\beta'_1)^2 - \gamma_2^2 - (\gamma'_2)^2 + 3 (\gamma'_3)^2 -  2 \delta_1^2\\
      0 &= 4\alpha_1^2 - 3\beta_1^2 -3 (\beta'_1)^2 +3 \gamma_2^2 +3 (\gamma'_2)^2 +3 (\gamma'_3)^2 + 2 \delta_1^2\\
      0 &= \beta_1^2 + (\beta'_1)^2 + 3\gamma_2^2 + 3(\gamma'_2)^2 + 3 (\gamma'_3)^2 -  2 \delta_1^2~.
    \end{aligned}
  \end{equation}
  There are two branches of solutions.  In both of them $\alpha_3 = \beta'_1=\gamma'_2=\gamma'_3=0$.
  \begin{enumerate}
  \item Letting $\xi_{1,2,3}=\pm 1$, the first branch is given by
    \begin{equation}
      \beta_1 = \xi_1 \sqrt3 \alpha_2 \qquad 
      \gamma_2 = \xi_2 \alpha_2 \qquad\text{and}\qquad
      \delta_1 = \xi_3 \sqrt3 \alpha_2~.
    \end{equation}
    
  \item Letting $\xi_{1,2}=\pm 1$, the second branch is given by $\gamma_2=0$ and in addition
    \begin{equation}
      \beta_1 = \xi_1 \sqrt2 \alpha_1 \qquad\text{and}\qquad
      \delta_1 = \xi_2 \alpha_1~.
    \end{equation}
  \end{enumerate}

\end{enumerate}

In summary, we have two kinds of backgrounds with this geometry:
\begin{enumerate}
\item For $\xi_{1,2,3}=\pm 1$:
  \begin{equation}
    \begin{aligned}[m]
      \fGa &= \alpha_2 d\vartheta^2\\
      \fG &= \alpha_2 \left(\xi_1 \sqrt3 \nu \wedge d\vartheta^1 - \xi_2 d\vartheta^{134}\right)\\
      \fH &= 0\\
      \fF &= \xi_3 \sqrt3 \alpha_2 \left(\nu \wedge d\vartheta^{234} + \sigma \wedge d\vartheta^1\right)~,
    \end{aligned}
  \end{equation}
  with curvatures
  \begin{equation}
    R_0 = - 2\alpha_2^2 g_0 \qquad\text{and}\qquad R_1 = \tfrac34 \alpha_2^2 g_1~;
  \end{equation}
\item and for $\xi_{1,2}=\pm 1$:
  \begin{equation}
    \begin{aligned}[m]
      \fGa &= \alpha_1 d\vartheta^1\\
      \fG &= \xi_1 \sqrt2 \alpha_1 \nu \wedge d\vartheta^1\\
      \fF &= \xi_2 \alpha_1 \left(\nu \wedge d\vartheta^{234} + \sigma \wedge d\vartheta^1\right)~,
    \end{aligned}
  \end{equation}
  with curvatures
  \begin{equation}
    R_0 = - \alpha_1^2 g_0 \qquad\text{and}\qquad R_1 = \half \alpha_1^2 g_1~.
  \end{equation}
  This latter branch is precisely (up to relabeling) the one we found earlier when looking for backgrounds with geometries \hyperlink{geom:AdS2S4S3T1}{$AdS_2\times S^4 \times S^3 \times T^1$} and \hyperlink{geom:AdS2S4S2T2}{$AdS_2\times S^4 \times S^2 \times T^2$}.
\end{enumerate}

Either of these two branches gives solutions for $\AdS_2 \times S^2 \times S^2 \times T^4$.

\subsubsection*{\texorpdfstring{$\mathbf{AdS_2 \times S^3 \times S^2 \times T^3}$}{AdS2 x S3 x S2 x T3}}\hypertarget{geom:AdS2S3S2T3}{}

This Ansatz has $(m_1,m_3,m_5^+) = (3,8,5)$, for a total of 24 parameters. The resultant system of polynomials does not lend itself to symbolic solution.  Nonetheless, we can exhibit an exact solution of the following form with $\xi_{1,2,3}=\pm 1$:
\begin{equation}  
  \begin{aligned}[m]  
   \fGa &= \kappa d\vartheta^1\\
   \fH &=\xi_1\sqrt{\tfrac{5}{2}}\kappa\nu_1 + \xi_2 \sqrt{\alpha^2 + \kappa^2} \nu_0 \wedge d\vartheta^1 + \alpha \nu_2 \wedge d\vartheta^1 + \xi_3 \tfrac{1}{\sqrt{2}} d\vartheta^{123}\\
   \fG &= \fF = 0~,
 \end{aligned}  
\end{equation}
with curvatures
\begin{equation}
  R_0 = -\tfrac{1}{4}(2\alpha^2 + 3\kappa^2) g_0\qquad R_1 = \kappa^2 g_1\qquad R_2 = \tfrac{1}{4}(2\alpha^2 - \kappa^2) g_2~,  
\end{equation} 
giving solutions for $AdS_2 \times S^3 \times S^2 \times T^3$, $AdS_2 \times S^3 \times T^5$ and $AdS_2 \times S^3 \times H^2 \times T^3$.  Another exact solution is given by the following, with $\xi_{1,2,3,4}=\pm 1$:
\begin{equation}  
  \begin{aligned}[m]  
   \fGa &= \kappa d\vartheta^3\\
   \fH &=\xi_1 \sqrt{\frac{43-\sqrt{57}}{28}} \kappa \nu_1 + \xi_2 \sqrt{\frac{9+5\sqrt{57}}{56}} \kappa \nu_0 \wedge d\vartheta^3 + \xi_3 \frac{4 \sqrt{3}}{3+\sqrt{57}}\kappa \nu_2 \wedge d\vartheta^3\\
   &- \xi_1 \xi_2 \xi_3 \sqrt{\frac{\sqrt{57}-3}{4}}\\
   \fG &= 0\\
   \fF &= \xi_4 \sqrt{\frac{2(6+\sqrt{57})}{7}} \kappa (\nu_1 \wedge d\vartheta^{12} + \nu_0 \wedge \nu_2 \wedge d\vartheta^3) +\xi_1 \xi_2 \xi_4 \tfrac{1}{\sqrt{2}}\kappa (\nu_0 \wedge d\vartheta^{123} + \nu_1 \wedge \nu_2)\\
   &+ \xi_1 \xi_3 \xi_4 \sqrt{2} \kappa (\nu_0 \wedge \nu_1 - \nu_2 \wedge d\vartheta^{123})~,
 \end{aligned}  
\end{equation}
with curvatures
\begin{equation}
  R_0 = -\frac{29 + \sqrt{57}}{16}\kappa^2 g_0\qquad R_1 = \kappa^2 g_1\qquad R_2 = \frac{13 + \sqrt{57}}{16}\kappa^2 g_2~,  
\end{equation}
giving a solution for $AdS_2 \times S^3 \times S^2 \times T^3$.

\subsubsection*{\texorpdfstring{$\mathbf{AdS_2 \times S^2 \times S^2 \times S^2 \times T^2}$}{AdS2 x S2 x S2 x S2 x T2}}\hypertarget{geom:AdS2S2S2S2T2}{}

This Ansatz has $(m_1,m_3,m_5^+) = (2,8,6)$, for a total of 24 parameters.  The resultant system of polynomials does not lend itself to symbolic solution.  However, we know that solutions exist for $AdS_2 \times S^2 \times S^2 \times S^2 \times T^2$ and $AdS_2 \times S^2 \times S^2 \times H^2 \times T^2$ as limits of the solutions for \hyperlink{geom:AdS2S4S2T2}{$AdS_2 \times S^4 \times S^2 \times T^2$} and \hyperlink{geom:AdS2S4S2T2}{$AdS_2 \times S^4 \times H^2 \times T^2$}.

\subsubsection*{\texorpdfstring{$\mathbf{AdS_2 \times \mathbb{CP}^2 \times T^4}$}{AdS2 x CP2 x T4}}\hypertarget{geom:AdS2CP2T4}{}

This Ansatz has $(m_1,m_3,m_5^+) = (4,12,10)$, for a total of 38 parameters.  The resultant system of polynomials does not lend itself to symbolic solution.  However, we know that solutions exist as a limit of \hyperlink{geom:AdS2CP2S3T1}{$AdS_2 \times \mathbb{CP}^2 \times S^3 \times T^1$}.

\subsubsection*{\texorpdfstring{$\mathbf{AdS_2 \times S^3 \times T^5}$}{AdS2 x S3 x T5}}\hypertarget{geom:AdS2S3T5}{}

This Ansatz has $(m_1,m_3,m_5^+) = (5,16,11)$, for a total of 48 parameters.  The resultant system of polynomials does not lend itself to symbolic solution.  However, we know that solutions exist as limits of \hyperlink{geom:AdS2S3S3T2}{$AdS_2 \times S^3 \times S^3 \times T^2$}, \hyperlink{geom:AdS2CP2S3T1}{$AdS_2 \times \CP^2 \times S^3 \times T^1$} and \hyperlink{geom:AdS2S3S2T3}{$AdS2 x S3 x S2 x T3$}.

\subsubsection*{\texorpdfstring{$\mathbf{AdS_2 \times S^2 \times S^2 \times T^4}$}{AdS2 x S2 x S2 x T4}}\hypertarget{geom:AdS2S2S2T4}{}

This Ansatz has $(m_1,m_3,m_5^+) = (4,16,12)$, for a total of 48 parameters.  The resultant system of polynomials does not lend itself to symbolic solution.  However, we know that solutions exist for both $AdS_2 \times S^2 \times S^2 \times T^4$ and $AdS_2 \times S^2 \times H^2 \times T^4$ as a limits of \hyperlink{geom:AdS2S3S2S2T1}{$AdS_2 \times S^3 \times S^2 \times S^2 \times T^1$} and \hyperlink{geom:AdS2S3S2S2T1}{$AdS_2 \times S^3 \times S^2 \times H^2 \times T^1$} respectively.

\subsubsection*{\texorpdfstring{$\mathbf{AdS_2 \times S^2 \times T^6}$}{AdS2 x S2 x T6}}\hypertarget{geom:AdS2S2T6}{}

This Ansatz has $(m_1,m_3,m_5^+) = (6,32,26)$, for a total of 96 parameters.  The resultant system of polynomials does not lend itself to symbolic solution.  However, we know that solutions exist as a limits of \hyperlink{geom:AdS2S5S2T1}{$AdS_2 \times S^5 \times S^2 \times T^1$}, \hyperlink{geom:AdS2S5S2T1}{$AdS_2 \times SLAG_3 \times S^2 \times T^1$}, \hyperlink{geom:AdS2S4S2T2}{$AdS_2 \times S^4 \times S^2 \times T^2$}, and \hyperlink{geom:AdS2S3H3S2}{$AdS_2 \times S^3 \times H^3 \times S^2$}.

\section{Summary}
\label{sec:summary}

We have identified (up to local isometry) all homogeneous backgrounds of type IIB supergravity where the underlying space is a ten-dimensional lorentzian symmetric space and in about two thirds of all cases have solved exactly for the moduli space.  There are two classes of solutions: those with underlying geometry a (possibly degenerate) Cahen--Wallach spaces and those with underlying geometry $\AdS_d \times K^{10-d}$ for $2 \leq d \leq 5$.  The latter class is summarised in two tables, depending on whether or not we have determined the exact moduli space.  In Table~\ref{tab:AdSModuli} we list those backgrounds for which we have determined fully the moduli space.  There are three numbers associated to each such background, corresponding to the dimension of sub-moduli spaces.  There are three types of moduli associated to such backgrounds.  Firstly, we have the \emph{geometric} moduli, corresponding to the free parameters in the given solutions.  One of these moduli always corresponds to the homothetic action of $\RR^+$ discussed in Section \ref{sec:homoth-invar-field}, hence the \emph{geometric} column is always $\geq 1$.  The \emph{duality} column is the dimension of the $SL(2,\RR)$-orbit of the background.  This number can be one of the following:
\begin{itemize}
\item[0:] this corresponds to backgrounds where $\fGa\neq 0$;
\item[2:] this corresponds to backgrounds where $\fGa=\fH=\fG=0$, so that the duality orbit is parametrised by the axi-dilaton $\tau$; and
\item[3:] this corresponds to backgrounds where $\fGa=0$ but $\fH$ (or $\fG$) are nonzero.
\end{itemize}
For the geometries \hyperlink{geom:AdS2CP3T2}{$AdS_2 \times G_{\mathbb{R}}^{+}(2,5) \times T^2$} and \hyperlink{geom:AdS2CP3T2}{$AdS_2 \times \mathbb{CP}^3 \times T^2$} there are two classes of branches with different number of duality moduli parameters: either 0 or 3.  Finally, the the third column, labelled \emph{other}, is the dimension of the generic orbit of the action of $SO(n)$ on backgrounds with geometries having a $T^n$ factor.  The only nonzero value among the geometries for which we have determined the full moduli space occurs for \hyperlink{geom:AdS3S3T4}{$AdS_3 \times S^3 \times T^4$}.  The moduli parametrise the orbit of $SO(4)$ acting on a nonzero self-dual (or anti-self-dual) 2-form in $\RR^4$. 

Table~\ref{tab:AdSUnknown} lists those geometries for which we have not been able to determine the full moduli space.  Such geometries have a status next to them on the table.  If the status says \emph{Some exact solutions}, it means that we have constructed some exact solutions, but have not managed to fully solve the equations and hence cannot claim to have determined the full moduli space.  Two of the possible statuses concern backgrounds whose existence can be deduced from other backgrounds.  In some cases such backgrounds exist as limits of other backgrounds when the radius of curvature of one of the riemannian factors goes to infinity.  These are indicated as \emph{$\exists$ as limit of} the relevant background.  Finally, there are backgrounds whose status is indicated as \emph{$\exists$ from}.  This can mean two things.  It can denote those backgrounds which were found by looking at geometries with fewer flat directions, but where the field equations forced one or more of the riemannian factors to be flat; or it can denote those backgrounds which are found by the second observation in Section \ref{sec:further-observations}.  In both cases we have families of solutions, but not necessarily the full moduli space.

There are no great surprises in the list of backgrounds.  In fact, the only backgrounds which are not $AdS$-sphere-flat products or plane waves are \hyperlink{geom:AdS2SU3}{$AdS_2 \times SU(3)$}, \hyperlink{geom:AdS2CP3T2}{$AdS_2 \times \mathbb{CP}^3 \times T^2$} and \hyperlink{geom:AdS2CP3T2}{$AdS_2 \times G_\RR^+(2,5) \times T^2$}.

The next step in this research programme is to identify which of these backgrounds are supersymmetric since, conjecturally\footnote{We thank Patrick Meessen for reminding us of this.}, they include all backgrounds preserving more than $\tfrac34$ of the supersymmetry.  Some backgrounds, such as those with $\mathbb{CP}^2$, $SLAG_3$, or $G_{\mathbb{R}}^{+}(2,5)$ factors are manifestly not spin --- although $\CP^2$ and $G_\RR^+(2,5)$ are spin${}^c$ --- and so they cannot be supersymmetric.

\begin{table}[h!]
  \centering
  \renewcommand{\arraystretch}{1.2}
  \caption{$AdS_d$ backgrounds with known moduli space}\label{tab:AdSModuli}
  \begin{tabular}{l|ccc}
    \multicolumn{1}{c|}{Geometry} & \multicolumn{3}{c}{Moduli}\\
    & geometric & duality & other\\\hline
    \hyperlink{geom:AdS5S5}{$AdS_5 \times S^5$} & 1 & 2 & 0\\
    \hyperlink{geom:AdS5S5}{$AdS_5 \times SLAG_3$} & 1 & 2 & 0\\
    \hyperlink{geom:AdS5S3S2}{$AdS_5 \times S^3 \times S^2$} & 1 & 2 & 0\\[10pt]
    \hyperlink{geom:AdS4S3S2T1}{$AdS_4 \times S^3 \times S^2 \times T^1$} & 1 & 0 & 0\\[10pt]
    \hyperlink{geom:AdS3S5H2}{$AdS_3 \times S^5 \times H^2$} & 1 & 2 & 0\\
    \hyperlink{geom:AdS3S5H2}{$AdS_3 \times SLAG_3 \times H^2$} & 1 & 2 & 0\\
    \hyperlink{geom:AdS3S3S2H2}{$AdS_3 \times S^3 \times S^2 \times H^2$} & 2 & 2 & 0\\
    \hyperlink{geom:AdS3S3S3T1}{$AdS_3 \times S^3 \times S^3 \times T^1$} & 2 & 3 & 0\\
    \hyperlink{geom:AdS3S3S2T2}{$AdS_3 \times S^3 \times S^2 \times T^2$} & 2 & 3 & 0\\
    \hyperlink{geom:AdS3S3T4}{$AdS_3 \times S^3 \times T^4$} & 2 & 3 & 2\\[10pt]
    \hyperlink{geom:AdS2SU3}{$AdS_2 \times SU(3) $} & 1 & 2 & 0\\
    \hyperlink{geom:AdS2CP3T2}{$AdS_2 \times G_{\mathbb{R}}^{+}(2,5) \times T^2$} & 1 & 0/3 & 0\\
    \hyperlink{geom:AdS2CP3T2}{$AdS_2 \times \mathbb{CP}^3 \times T^2$} & 1 & 0/3 & 0\\
    \hyperlink{geom:AdS2S5H3}{$AdS_2 \times S^5 \times H^3 $} & 1 & 2 & 0\\
    \hyperlink{geom:AdS2S5H3}{$AdS_2 \times SLAG_3 \times H^3 $} & 1 & 2 & 0\\
    \hyperlink{geom:AdS2S4S3T1}{$AdS_2 \times H^4 \times S^3 \times T^1 $} & 1 & 0 & 0\\
    \hyperlink{geom:AdS2S5S2T1}{$AdS_2 \times S^5 \times S^2 \times T^1 $} & 2 & 0 & 0\\
    \hyperlink{geom:AdS2S5S2T1}{$AdS_2 \times SLAG_3 \times S^2 \times T^1 $} & 2 & 0 & 0\\
    \hyperlink{geom:AdS2S3H3S2}{$AdS_2 \times S^3 \times H^3 \times S^2 $} & 2 & 2 & 0\\
    \hyperlink{geom:AdS2S3S3T2}{$AdS_2 \times S^3 \times S^3 \times T^2 $} & 2 & 3 & 0\\
    \hyperlink{geom:AdS2CP2S3T1}{$AdS_2 \times \mathbb{CP}^2 \times S^3 \times T^1 $} & 2 & 0 & 0\\
    \hyperlink{geom:AdS2CP2S3T1}{$AdS_2 \times \mathbb{CH}^2 \times S^3 \times T^1 $} & 2 & 0 & 0\\
    \hyperlink{geom:AdS2S4S2T2}{$AdS_2 \times S^4 \times S^2 \times T^2$} & 2 & 0 & 0\\
    \hyperlink{geom:AdS2S4S2T2}{$AdS_2 \times S^4 \times H^2 \times T^2$} & 2 & 0 & 0\\
    \hyperlink{geom:AdS2S4T4}{$AdS_2 \times S^4 \times T^4$} & 2 & 0 & 0
  \end{tabular}
\end{table}

\begin{table}[h!]
  \centering
  \renewcommand{\arraystretch}{1.2}
  \caption{$AdS$ backgrounds with unknown moduli space}\label{tab:AdSUnknown}
  \begin{tabular}{>{$}l<{$}|l}
    \multicolumn{1}{c|}{Geometry} & Status\\
    \hline
    AdS_3 \times S^2 \times S^2 \times T^3 & \hyperlink{geom:AdS3S2S2T3}{Some exact solutions}\\
    AdS_3 \times S^2 \times T^5 & $\exists$ as limit of \hyperlink{geom:AdS3S3S2T2}{$AdS_3 \times S^3 \times S^2 \times T^2$}\\[10pt]
    AdS_2 \times S^3 \times S^2 \times S^2 \times T^1 & \hyperlink{geom:AdS2S3S2S2T1}{Some exact solutions}\\
    AdS_2 \times S^3 \times H^2 \times S^2 \times T^1 & \hyperlink{geom:AdS2S3S2S2T1}{Some exact solutions}\\
    AdS_2 \times S^3 \times H^2 \times H^2 \times T^1 & \hyperlink{geom:AdS2S3S2S2T1}{Some exact solutions}\\
    AdS_2 \times \mathbb{CP}^2 \times S^2 \times T^2 & \hyperlink{geom:AdS2CP2S2T2}{Some exact solutions}\\
    AdS_2 \times \mathbb{CP}^2 \times H^2 \times T^2 & \hyperlink{geom:AdS2CP2S2T2}{Some exact solutions}\\
    AdS_2 \times S^3 \times S^2 \times T^3 & \hyperlink{geom:AdS2S3S2T3}{Some exact solutions}\\
    AdS_2 \times S^3 \times H^2 \times T^3 & \hyperlink{geom:AdS2S3S2T3}{Some exact solutions}\\
    AdS_2 \times S^2 \times S^2 \times S^2 \times T^2 & $\exists$ from \hyperlink{geom:AdS2S4S2T2}{$AdS_2 \times S^4\times S^2 \times T^2$}\\
    AdS_2 \times S^2 \times S^2 \times H^2 \times T^2 & $\exists$ from \hyperlink{geom:AdS2S4S2T2}{$AdS_2 \times S^4\times H^2 \times T^2$}\\
    AdS_2 \times \mathbb{CP}^2 \times T^4 & $\exists$ from \hyperlink{geom:AdS2CP2S3T1}{$AdS_2 \times \mathbb{CP}^2 \times S^3 \times T^1$}\\
    AdS_2 \times S^3 \times T^5 & $\exists$ as limit of \hyperlink{geom:AdS2S3S3T2}{$AdS_2 \times S^3 \times S^3 \times T^2$}, \hyperlink{geom:AdS2CP2S3T1}{$AdS_2 \times \mathbb{CP}^2 \times S^3 \times T^1$},\\
    & \hphantom{$\exists$}\hyperlink{geom:AdS2S3S2T3}{$AdS_2 \times S^3 \times S^2 \times T^3$} and 
\hyperlink{geom:AdS2S3S2T3}{$AdS_2 \times S^3 \times H^2 \times T^3$}\\
    AdS_2 \times S^2 \times S^2 \times T^4 & $\exists$ from \hyperlink{geom:AdS2S3S2S2T1}{$AdS_2 \times S^3 \times S^2 \times S^2 \times T^1$} and \hyperlink{geom:AdS2S4T4}{$AdS_2 \times S^4 \times T^4$}\\
    AdS_2 \times S^2 \times H^2 \times T^4 & $\exists$ from \hyperlink{geom:AdS2S3S2S2T1}{$AdS_2 \times S^3 \times S^2 \times H^2 \times T^1$}\\
    AdS_2 \times S^2 \times T^6 & $\exists$ from \hyperlink{geom:AdS2S5S2T1}{$AdS_2 \times S^5 \times S^2 \times T^1$}, \hyperlink{geom:AdS2S5S2T1}{$AdS_2 \times SLAG_3 \times S^2 \times T^1$}\\
    & \hphantom{$\exists$} and \hyperlink{geom:AdS2S4S2T2}{$AdS_2 \times S^4 \times S^2 \times T^2$}; $\exists$ as limit of \hyperlink{geom:AdS2S3H3S2}{$AdS_2 \times S^3 \times H^3 \times S^2$}
  \end{tabular}
\end{table}

\appendix

\section{Inadmissible geometries}
\label{sec:inadm-geom}

In this appendix we list those geometries not already ruled out by general arguments but which we have shown do not admit any solutions.  Although we list the geometries by using the compact versions of the riemannian symmetric spaces which appear, their noncompact duals are similarly ruled out.

\begin{multicols}{3}
  \begin{itemize}
  \item $AdS_6 \times S^2 \times T^2$
  \item $AdS_5 \times S^4 \times T^1$
  \item $AdS_5 \times S^3 \times T^2$
  \item $AdS_5 \times \CP^2 \times T^1$
  \item $AdS_5 \times S^2 \times S^2 \times T^1$
  \item $AdS_5 \times S^2 \times T^3$
  \item $AdS_4 \times S^5 \times T^1$
  \item $AdS_4 \times SLAG_3 \times T^1$
  \item $AdS_4 \times S^4 \times T^2$
  \item $AdS_4 \times S^3 \times S^3$
  \item $AdS_4 \times S^3 \times T^3$
  \item $AdS_4 \times \CP^2 \times T^2$
  \item $AdS_4 \times S^2 \times S^2 \times T^2$
  \item $AdS_4 \times S^2 \times T^4$
  \item $AdS_3 \times S^5 \times T^2$
  \item $AdS_3 \times SLAG_3 \times T^2$
  \item $AdS_3 \times S^4 \times S^3$
  \item $AdS_3 \times \CP^2 \times S^3$
  \item $AdS_3 \times S^4 \times S^2 \times T^1$
  \item $AdS_3 \times G_\RR^+(2,5) \times T^1$
  \item $AdS_3 \times \CP^3 \times T^1$
  \item $AdS_3 \times S^4 \times T^3$
  \item $AdS_3 \times \CP^2 \times S^2 \times T^1$
  \item $AdS_3 \times \mathbb{CP}^2 \times T^3$
  \item $AdS_3 \times S^2 \times S^2 \times S^2 \times T^1$
  \item $AdS_2 \times S^6 \times T^2$
  \item $AdS_2 \times S^5 \times T^3$
  \item $AdS_2 \times SLAG_3 \times T^3$
  \end{itemize}
\end{multicols}

In most cases the geometry is ruled out by analysing the Einstein equation along the flat directions and showing that their flatness forces the vanishing of all the parameters in the Ansatz, which contradicts the fact that these geometries are not Ricci-flat.  There are three geometries which require other arguments:  $AdS_4 \times S^3 \times S^3$ is ruled out because the basis for invariant 3-forms consists of two 3-forms belonging to different riemannian factors and hence they cannot simultaneously satisfy the second and third equations in \eqref{eq:IIBEoMsym};$AdS_3 \times \mathbb{CP}^2 \times T^3$, $AdS_2 \times S^5 \times T^3$ and $AdS_2 \times SLAG_3 \times T^3$ turn out to be trickier and the proofs of their inadmissibility appear below.

\subsection{\texorpdfstring{$AdS_2 \times S^5 \times T^3$ and $AdS_2 \times SLAG_3 \times T^3$}{AdS2 x S5 x T3 and AdS2 x SLAG3 x T3}}

The invariant forms are the volume form $\nu$ of $AdS_2$, the volume form $\sigma$ of $S^5$ (or $SLAG_3$) and any constant-coefficient form on $T^3$.  Let $d\vartheta^i$, $i=1,2,3$, denote an orthonormal coframe on $T^3$ and let $\tau =  d\vartheta^{123}$ be the corresponding volume form.  The most general Ansatz for the forms in this geometry, taking into account the self-duality of $\fF$, is
\begin{equation}
  \begin{aligned}[m]
    \fGa &= \alpha\\
    \fG &= \nu \wedge \beta + \gamma \tau\\
    \fH &= \nu \wedge \beta' + \gamma' \tau\\
    \fF &= \delta (\nu \wedge \tau + \sigma)~,
  \end{aligned}
\end{equation}
where $\alpha,\beta,\beta'$ are invariant 1-forms on $T^3$ and $\gamma,\gamma',\delta$ are constants.  The field equations \eqref{eq:IIBEoMsym} in this Ansatz become
\begin{equation}
  \begin{aligned}[m]
    -|\beta'|^2 + (\gamma')^2 &= -|\beta|^2 + \gamma^2 + 2 |\alpha|^2\\
    0 &= -\left<\beta,\beta'\right> + \gamma\gamma'\\
    0 &= \delta (\nu \wedge \beta' + \gamma' \tau)\wedge \sigma\\
    0 &= \delta (\nu \wedge \beta + \gamma \tau)\wedge \sigma\\
  \end{aligned}
\end{equation}
and in addition the Einstein equation.  (The equation $\fG \wedge \fH = 0$ is identically satisfied in this Ansatz.)  The last two equations say that either $\fF=0$ or else $\fG=\fH=0$.  This gives rise to two branches.

\begin{enumerate}
\item First branch: $\fF\neq 0$, so that $\fG = \fH = 0$.  The first of the above field equations then says that $\fGa=0$ as well.  The Einstein equations along $T^3$ then become
  \begin{equation}
    0 = -\tfrac14 \delta^2 \delta_{ij}
  \end{equation}
  which implies that $\fF=0$, contradicting the hypothesis.

\item Second branch: $\fF=0$.  Using the $\SO(3)$ symmetry of the $T^3$ metric, we may rotate $\beta$ so that $\beta = \beta_1 d\vartheta^1$ and then use the stabilising $\SO(2)$ to rotate $\beta' = \beta'_1 d\vartheta^1 + \beta'_2 d\vartheta^2$.  Let us then consider the $(33)$ component of the Einstein equation along $T^3$:
  \begin{equation}
    0 = \half (\alpha_3)^2 + \tfrac18 \left(3 \gamma^2 + 3 (\gamma')^2 + |\beta|^2 + |\beta'|^2 \right)~,
  \end{equation}
  whence, in particular, $\fF = \fH = 0$, but then the first of the above field equations imply that $\alpha = 0$ and hence all forms are zero, which would imply that the spacetime is Ricci-flat, which is absurd.
\end{enumerate}

\subsection{\texorpdfstring{$AdS_3 \times \mathbb{CP}^2 \times T^3$}{AdS3 x CP2 x T3}}

The invariant forms are the volume form $\nu$ for $AdS_3$, powers of the Kähler form $\omega$ of $\CP^2$ and any constant-coefficient form on $T^3$.  We will let $d\vartheta^i$, $i=1,2,3$, denote an orthonormal coframe for $T^3$ and $\tau = d\vartheta^{123}$ the corresponding volume form.  The volume form on $\CP^2$ is $\half \omega^2$, whence $|\omega|^2 = 2$.  The most general Ansatz in this geometry, taking into account the self duality of $\fF$, is
\begin{equation}
  \begin{aligned}[m]
    \fGa &= \alpha\\
    \fG &= f_0 \nu + f_1 \tau + \omega \wedge \beta\\
    \fH &= f'_0 \nu + f'_1 \tau + \omega \wedge \beta'\\
    \fF &= f_2 (\nu - \tau) \wedge \omega + \nu\wedge \star \gamma - \half \omega^2 \wedge \gamma~,
  \end{aligned}
\end{equation}
where $\alpha,\beta,\beta',\gamma$ are constant-coefficient 1-forms on $T^3$ and $f_0,f'_0,f_1,f'_1,f_2$ are constants.  The field equations \eqref{eq:IIBEoMsym} in this Ansatz become
\begin{equation}
  \label{eq:MaxEqnA2}
  \begin{aligned}[m]
    -(f'_0)^2 + (f'_1)^2 + 2 |\beta'|^2 &= -f_0^2 + f_1^2 + 2|\beta|^2 + 2 |\alpha|^2\\
    0 &= -f_0 f'_0 + f_1 f'_1 + 2 \left<\beta,\beta'\right>\\
    0 &= f_2(f_0+f_1) + \left<\beta,\gamma\right>\\
    0 &= f_2 \beta + \half f_0 \gamma\\
    0 &= f_2(f'_0+f'_1) + \left<\beta',\gamma\right>\\
    0 &= f_2 \beta' + \half f'_0 \gamma\\
    0 &= f_0 f'_1 - f_1 f'_0\\
    0 &= f_0 \beta' - f'_0 \beta\\
    0 &= \beta \wedge \beta'~,
  \end{aligned}
\end{equation}
together with the Einstein equation.  We shall only need the components of the Einstein equation along $T^3$, which are given by
\begin{multline}
  \label{eq:T3EEA2}
  0 = \half \alpha_i\alpha_j + \beta_i \beta_j + \beta'_i\beta'_j + \half \gamma_i\gamma_j \\
  + \delta_{ij} \left( \tfrac18 (f_0^2 + (f'_0)^2) + \tfrac38 (f_1^2 + (f'_1)^2) + \half f_2^2 - \tfrac14 (|\beta|^2 + |\beta'|^2 + |\gamma|^2) \right)~.
\end{multline}

The second and third equations from the bottom in \eqref{eq:MaxEqnA2} imply that $\fG$ and $\fH$ satisfy $f'_0 \fG - f_0 \fH = 0$.  Let us first assume that at least one of $f_0$ and $f'_0$ is different from zero.  In that case, $\fG$ and $\fH$ are collinear.  Since $\fG$ and $\fH$ are perpendicular, two situations can occur: either at least one of them vanishes or else, if both are nonvanishing, they have zero norm, in which case $\alpha=0$ from the first of the above equations.  We therefore have three branches to consider:
\begin{enumerate}
\item First branch: $\fG\neq 0 \neq \fH$, hence $\alpha = 0$.  In this case, $\fH = c \fG$ for some constant $c$.  We may use the $\SO(3)$ symmetry of the $T^3$ metric to set $\beta = \beta_1 d\vartheta^1$ and use the stabilising $\SO(2)$ to set $\gamma = \gamma_1 d\vartheta^1 + \gamma_2 d\vartheta_2$.  In particular, the $(33)$ component of equation~\eqref{eq:T3EEA2} says that the term multiplying $\delta_{ij}$ vanishes, whence equation ~\eqref{eq:T3EEA2} becomes
  \begin{equation}
    0 = (c^2+1)\beta_i \beta_j + \half \gamma_i\gamma_j~.
  \end{equation}
Tracing with $\delta^{ij}$, we find
\begin{equation}
  0 = (c^2 + 1)|\beta|^2 + \half |\gamma|^2~,
\end{equation}
which says that $\beta = \gamma = 0$.  This renders the term multiplying $\delta_{ij}$ in the Einstein equation a sum of non-negative terms, whence its vanishing
\begin{equation}
  \tfrac18 (c^2+1) f_0^2 + \tfrac38 (c^2+1) f_1^2 + \half f_2^2 = 0
\end{equation}
imposes $f_0=f_1=f_2=0$, contradicting the fact that $\fG\neq0$.

\item Second branch: $\fH = 0$.  Here $f'_0=f'_1=\beta'=0$.  We recognise two sub-branches depending on whether or not $f_0$ vanishes.
  \begin{enumerate}
  \item First sub-branch: $f_0 \neq 0$.  Then $\gamma = -2\frac{f_2}{f_0}\beta$.  We may use the $\SO(3)$ symmetry to set $\alpha = \alpha_1 d\vartheta^1$ and the stabilising $\SO(2)$ to set $\beta = \beta_1 d\vartheta^1 + \beta_2 d\vartheta^2$.  The $(33)$ component of the Einstein equation \eqref{eq:T3EEA2} then sets the term multiplying $\delta_{ij}$ to zero, whence the equation becomes
    \begin{equation}
      0 = \half \alpha_i\alpha_j + \beta_i \beta_j + \half \gamma_i\gamma_j 
    \end{equation}
    which upon tracing says that $\alpha = \beta = \gamma = 0$.  Back into the term multiplying $\delta_{ij}$ in the Einstein equation, we arrive at
    \begin{equation}
      0 = \tfrac18 f_0^2 + \tfrac38 f_1^2 + \half f_2^2~,
    \end{equation}
    whence $f_0 = f_1 = f_2 = 0$, contradicting the hypothesis.

  \item Second sub-branch: $f_0 = 0$.  The first of the equations in \eqref{eq:MaxEqnA2} the forces $f_1 = \alpha = \beta = 0$.  The Einstein equation \eqref{eq:T3EEA2} then becomes
    \begin{equation}
      0 = \half \gamma_i\gamma_j + \delta_{ij} \left(\half f_2^2 - \tfrac14 |\gamma|^2\right)~.
    \end{equation}
    We can use the $\SO(3)$ symmetry to set $\gamma =\gamma_1 d\vartheta^1$, whence the $(33)$ component of the above equation sets $f_2^2 = \half |\gamma|^2$ and hence $\gamma_i\gamma_j = 0$.  Tracing this equation sets $\gamma=0$ and hence $f_2=0$.  Hence all the forms vanish, imposing Ricci-flatness of the geometry, which is absurd.
  \end{enumerate}

\item Third branch: $\fG = 0$.  Here $f_0 = f_1 = \beta = 0$ and we proceed along the same route as in the previous branch.  We recognise two branches according to whether $f'_0$ does or does not vanish.
  \begin{enumerate}
  \item First sub-branch: $f'_0 \neq 0$.  The argument is \emph{mutatis mutandis} the same as in the first sub-branch of the previous branch.  We arrive at $f'_0=0$ which contradicts the hypothesis.
  \item Second sub-branch: $f'_0 = 0$.  Then the fourth equation from the bottom in \eqref{eq:MaxEqnA2} says that $f_2 \beta' = 0$ and multiplying the middle equation in \eqref{eq:MaxEqnA2} by $f_2$ we also find that $f_2f'_1=0$.  If $f_2\neq 0$ then $\beta'=f'_1=0$ whence $\fH = 0$ and we are back to the second branch.  So let us take $f_2=0$.  Then the remaining equations are
    \begin{equation}
      \label{eq:RemEqnsA2}
      \begin{aligned}[m]
        0 &= \left<\beta',\gamma\right>\\
        0 &= (f'_1)^2 + 2|\beta'|^2 - 2|\alpha|^2\\
        0 &= \half \alpha_i\alpha_j + \beta'_i\beta'_j + \half \gamma_i\gamma_j + \tfrac14 \delta_{ij} \left(\tfrac32 (f'_1)^2 - |\beta'|^2 - |\gamma|^2\right)~.
      \end{aligned}
    \end{equation}
    Tracing the last of the above equations we arrive at
    \begin{equation}
      |\alpha|^2 + \half |\beta'|^2 + \tfrac94 (f'_1)^2  = \half |\gamma|^2
    \end{equation}
    which together with the second of the above equations allows us to rewrite the Einstein equation as
    \begin{equation}
      \label{eq:EEtoo}
      0 = \half \alpha_i\alpha_j + \beta'_i\beta'_j + \half \gamma_i\gamma_j - \delta_{ij} \left((f'_1)^2 + |\beta'|^2 \right)~.
    \end{equation}
    We now use the $\SO(3)$ symmetry to set $\beta' = \beta'_1 d\vartheta^1$ and then because $\left<\gamma,\beta'\right> = 0$, we may use the stabilising $\SO(2)$ symmetry to set $\gamma = \gamma_2 d\vartheta^2$.  (This argument assumes implicitly that $\beta'\neq 0$.  The result is of course still valid: if $\beta'=0$ then we simply use the $\SO(3)$ symmetry to set $\gamma$ to the same expression.)  The $(13)$ and $(23)$ components of the Einstein equation \eqref{eq:EEtoo} say that $\alpha_1\alpha_3=0=\alpha_2\alpha_3$, whereas the $(33)$ component says that $\half \alpha_3^2 = (f'_1)^2 + |\beta'|^2$.  If $\alpha_3=0$ then $f'_1=\beta' = 0$, so $\alpha = \gamma = 0$ as well by tracing the Einstein equation.  This means that the geometry is forced to be Ricci-flat, which is absurd.  Hence we take $\alpha_3 \neq 0$, whence $\alpha_1=\alpha_2 =0$.  Hence we have that $\alpha$, $\beta'$ and $\gamma$ have components in different orthonormal directions.  The $(11)$ component of equation \eqref{eq:EEtoo} says that $(\beta'_1)^2 = (f'_1)^2 + (\beta'_1)^2$, whence $f'_1 = 0$.  The $(33)$ component says that $\half \alpha_3^2 = (\beta'_1)^2$, whereas the middle equation in \eqref{eq:RemEqnsA2} says that $\alpha_3^2 = (\beta'_1)^2$, which says that $\alpha = \beta'=0$ and then the $(22)$ component of equation \eqref{eq:EEtoo} sets $\gamma=0$ as well.  Therefore all forms vanish contradicting the fact that this geometry is not Ricci-flat.
  \end{enumerate}
\end{enumerate}

Now let us deal with the case where both $f_0=f'_0=0$.  Multiplying the third equation in \eqref{eq:MaxEqnA2} by $f_2$ and using the fourth equation, which now reads $f_2\beta = 0$, we see that $f_2f_1 = 0$ and hence $\left<\beta,\gamma\right>=0$.  Doing the same with the fifth equation, we see that $f_2f'_1=0$ and hence that $\left<\beta',\gamma\right>=0$.  If $f_2 \neq0$ then $f_1=f'_1=\beta=\beta'=0$, whence $\fG=\fH=0$ and we are back the cases treated above.  Therefore let us take $f_2 = 0$.  The remaining equations are now given by
\begin{equation}
  \label{eq:MaxEqnA2last}
  \begin{aligned}[m]
    (f'_1)^2 + 2 |\beta'|^2 &= f_1^2 + 2|\beta|^2 + 2 |\alpha|^2\\
    0 &= f_1 f'_1 + 2 \left<\beta,\beta'\right>\\
    0 &= \left<\beta,\gamma\right>\\
    0 &= \left<\beta',\gamma\right>\\
    0 &= \beta \wedge \beta'~,
  \end{aligned}
\end{equation}
and the $T^3$ components of the Einstein equation are
\begin{equation}
  \label{eq:T3EEA2last}
 0 = \half \alpha_i\alpha_j + \beta_i \beta_j + \beta'_i\beta'_j + \half \gamma_i\gamma_j + \delta_{ij} \left( \tfrac38 (f_1^2 + (f'_1)^2) - \tfrac14 (|\beta|^2 + |\beta'|^2 + |\gamma|^2) \right)~.
\end{equation}
Tracing this equation we find
\begin{equation}
  0 = \half |\alpha|^2 + \tfrac14 |\beta|^2 + \tfrac14 |\beta'|^2 -\tfrac14|\gamma|^2 + \tfrac98 (f_1^2 + (f'_1)^2)~,
\end{equation}
whence $\gamma\neq0$ else all forms are forced to vanish, contradicting the fact that the geometry is not Ricci-flat.  Since $\gamma\neq 0$, the third and fourth equations in \eqref{eq:MaxEqnA2last} say that $\beta,\beta'$ lie on the same plane (namely, the plane perpendicular to $\gamma$) and the last equation in \eqref{eq:MaxEqnA2last} says that $\beta$ and $\beta'$ are actually collinear.  Let us use the $SO(3)$ symmetry of $T^3$ in order to set $\beta = \beta_1 d\vartheta^1$ and hence $\beta' = \beta'_1 d\vartheta^1$.  We then use the residual $SO(2)$ symmetry to set $\gamma = \gamma_2 d\vartheta^2$.  The $(13)$ and $(23)$ components of the Einstein equation \eqref{eq:T3EEA2last} then say that $\alpha_1\alpha_3=0$ and $\alpha_2\alpha_3=0$.  We branch according to whether $\alpha_3$ does or does not vanish.
\begin{enumerate}
\item $\alpha_3=0$.  Then the $(33)$ component of equation \eqref{eq:T3EEA2last} says that the terms multiplying $\delta_{ij}$ vanish, whence so do separately the other terms
  \begin{equation}
    \half \alpha_i\alpha_j + \beta_i \beta_j + \beta'_i\beta'_j + \half \gamma_i\gamma_j = 0~.
  \end{equation}
  Tracing we find that $\alpha = \beta = \beta' = \gamma = 0$ which then brings us to the case of $\fG$ and $\fH$ collinear, which was treated already.

\item $\alpha_3 \neq 0$, whence $\alpha_1 = \alpha_2 = 0$.  Then the $(11)$ and $(22)$ components of equation \eqref{eq:T3EEA2last} become
  \begin{equation}
    \begin{aligned}[m]
      \tfrac34 (\beta_1^2 + (\beta'_1)^2) + \tfrac38 (f_1^2 + (f'_1)^2) &= \tfrac14 \gamma_2^2\\
      \tfrac14 \gamma_2^2 + \tfrac38 (f_1^2 + (f'_1)^2) &= \tfrac14 (\beta_1^2 + (\beta'_1)^2)~.
    \end{aligned}
  \end{equation}
  Three times the second equation into the first gives that $\gamma_2=f_1=f'_1 =0$, contradicting that $\gamma\neq 0$.
\end{enumerate}

\bibliographystyle{utphys}
\bibliography{AdS,AdS3,ESYM,Sugra,Geometry}

\end{document}